\pdfoutput=1
\documentclass[5p,twocolumn]{elsarticle}

\usepackage{lineno,hyperref}
\modulolinenumbers[5]

\usepackage{amssymb,amsfonts,amsmath}
\usepackage{graphicx}
\usepackage{color}

\newcommand*{\vect}[1]{\boldsymbol{#1}}

\journal{Journal of Molecular Biology}

\biboptions{sort&compress,comma}

\begin{document}

\begin{frontmatter}

\title{Propensity to form amyloid fibrils is encoded as excitations in the free energy landscape of monomeric proteins}


\author[IPST]{Pavel I. Zhuravlev}
\author[IIS]{Govardhan Reddy}
\author[BU]{John E. Straub}

\author[IPST]{D.Thirumalai\corref{mycorrespondingauthor}}
\cortext[mycorrespondingauthor]{Corresponding author}
\ead{thirum@umd.edu}

\address[IPST]{Biophysics Program, Institute for Physical Science and Technology, Department of Chemistry and Biochemistry, University of Maryland, College Park, MD 20742, USA}
\address[IIS]{Solid State and Structural Chemistry Unit, Indian Institute of Science, Bangalore, 560 012, India}
\address[BU]{Department of Chemistry, Boston University, 590 Commonwealth Ave, Boston, MA, 02215-2521, USA}

\begin{abstract}
Protein aggregation, linked to many of diseases, is initiated when monomers access rogue conformations that are poised to form amyloid fibrils. We show, using simulations of src SH3 domain, that mechanical force enhances the population of the aggregation prone ($N^*$) states, which are rarely populated under force free native conditions, but are encoded in the spectrum of native fluctuations. The folding phase diagrams of SH3 as a function of denaturant concentration ($[C]$), mechanical force ($f$), and temperature exhibit an apparent two-state behavior, without revealing the presence of the elusive $N^*$ states. Interestingly, the phase boundaries separating the folded and unfolded states at all [C] and $f$ fall on a master curve, which can can be quantitatively described using an analogy to superconductors in a magnetic field. The free energy profiles as a function of the molecular extension ($R$), which are accessible in pulling experiments, ($R$), reveal the presence of a native-like $N^*$ with a disordered solvent-exposed amino terminal $\beta$-strand. The structure of the $N^*$ state is identical to that found in Fyn SH3 by NMR dispersion experiments. We show that the time scale for fibril formation can be estimated from the population of the $N^*$ state, determined by the free energy gap separating the native structure and the $N^*$ state, a finding that can be used to assess fibril forming tendencies of proteins. The structures of the $N^*$ state are used to show that oligomer formation and likely route to fibrils occur by a domain-swap mechanism in SH3 domain.
\end{abstract}


\end{frontmatter}


\section{Introduction}

The continuing effort to understand how proteins fold \cite{Schuler:2008ib,LindorffLarsen:2011gl,Onuchic:2004p2023,Shakhnovich:2006gs,Thirumalai:2010ia} is amply justified by the irrefutable link between misfolding\cite{Shea:2012wv,Straub:2010cd}, aggregation, and a growing list of diseases\cite{Aguzzi:2010gd,Chiti:2006fz}. It is now firmly established that all proteins, regardless of their role in causing diseases, form amyloid fibrils  under suitable conditions\cite{Bucciantini:2002it,Chiti:2006fz}. In most instances, newly synthesized proteins do not aggregate, but rather fold, carry out the intended functions, and are subsequently degraded. The potential deleterious effects of interactions between misfolded structures leading to fibrils  have made it urgent to understand the characteristics of proteins that harbor propensities to aggregate. Based on high resolution crystal structures of fibrils of small peptides, sequence-based methods have been introduced to identify motifs that harbor amyloidogenic tendencies\cite{Goldschmidt:2010dq,DeSimone:2011gt,Tartaglia:2008ie}. In addition to sequence, it is likely that the structures sampled by the monomer under native conditions encode not only the structures in the fibril state, but also the rate of fibril formation. Thus, in order to decipher the aggregation-prone states, a complete structural characterization of not only the native state, but also of higher free energy excitations, that are amyloidogenic, is needed.

Since the cascade of events driving a monomer to a fibril depends both on the protein sequence and external conditions, there are multiple scenarios within the standard nucleation growth theory of protein aggregation. Regardless of the scenario, we theorize that protein aggregation is initiated if the aggregation prone ensemble of structures (denoted as the $N^*$ state) is transiently populated due to thermal fluctuations or denaturation stress\cite{Thirumalai:2003wn,Tarus:2006gb,Massi:2001ww}. Thus, it is likely that both the tendency of a protein to form aggregates, and, more interestingly, the rate of fibril formation are determined by the population of the $N^*$ state. It also implies that the excitations in the spectrum of monomer conformations themselves are harbingers of protein aggregation. Consequently, identifying and revealing the stability and structure of the $N^*$ state should yield quantitative insights into protein aggregation.

Because $N^*$ is typically an excitation around the lowest free energy state (for a vast majority of proteins with a possible exception of mammalian prions\cite{Thirumalai:2003wn}), it is only sparsely populated, and thus is hard to detect experimentally. High structural and temporal resolutions are required to characterize structure and the extent of population of the $N^*$ state. Techniques like hydrogen exchange\cite{Hu:2013ck} and NMR spectroscopy\cite{Neudecker:2012kn} are successful in these endeavors. Here, we show that the elusive $N^*$ state can be more readily identified using single molecule force spectroscopy (SMFS)\cite{Jagannathan:uw,Jagannathan:2013bn}, because application of mechanical force ($f$) can enhance the population of high free energy states and slow down the dynamics. The only accessible experimental observable in the SMFS is the time-dependent change in the end-to-end distance ($R$), from which the $f$-dependent free energy profile can be determined. We show  that $F(R)$ can reveal the presence of the elusive $N^*$ state, the population of which is enhanced by the application of force ($R$ being the variable conjugate to $f$), thus establishing the potential applicability of the SMFS to characterize excited states in the free energy spectra of monomers.  

To reveal the nature of the $N^*$ ensemble, we performed simulations of a coarse-grained model of the SH3 domain from  the \textit{G.Gallus} src tyrosine kinase (PDB code 1SRL) at various values of constant force applied to the ends of the protein. We determined the folding phase diagrams in the $(f,T)$ and $(f,[C])$ planes ($[C]$ is the concentration of the denaturant guanidinium chloride (GdmCl)). The phase diagrams show that SH3 folding (triggered by changing $f$, $[C]$, or $T$) can be approximately characterized by a two-state model. The phase boundaries in the $(f,T)$ and $(f,[C])$ planes collapse onto a universal curve, which we quantify using an analogy to superconductors in a magnetic field. The $f$-dependent $F(R)$s  reveal the presence of a state $N^*$, which becomes prominent as $f$ increases. Structurally, the $N^*$ state corresponds to melting of the $\beta$-sheet formed by the N- and C-terminal strands. Surprisingly, an identical structure, with very low population, which has the same native fold as src SH3, has been recently identified in Fyn SH3 \cite{Neudecker:2012kn}. Force-unfolding kinetics shows that $N^*$ state becomes populated ahead of global unfolding. Most importantly, we establish that the structure of $N^*$ allows us to determine the mechanism of oligomer formation, which in turn naturally suggests a route to fibril formation. For the src SH3 protein this occurs by a domain-swap mechanism. Our work also shows that estimates of fibril formation times can be made using the population of $N^*$, thus establishing a direct link between the entire folding landscape and the propensity to aggregate.

\section{Results}

\begin{figure}[tbh]
	\centering
 		\includegraphics[width=8.7cm]{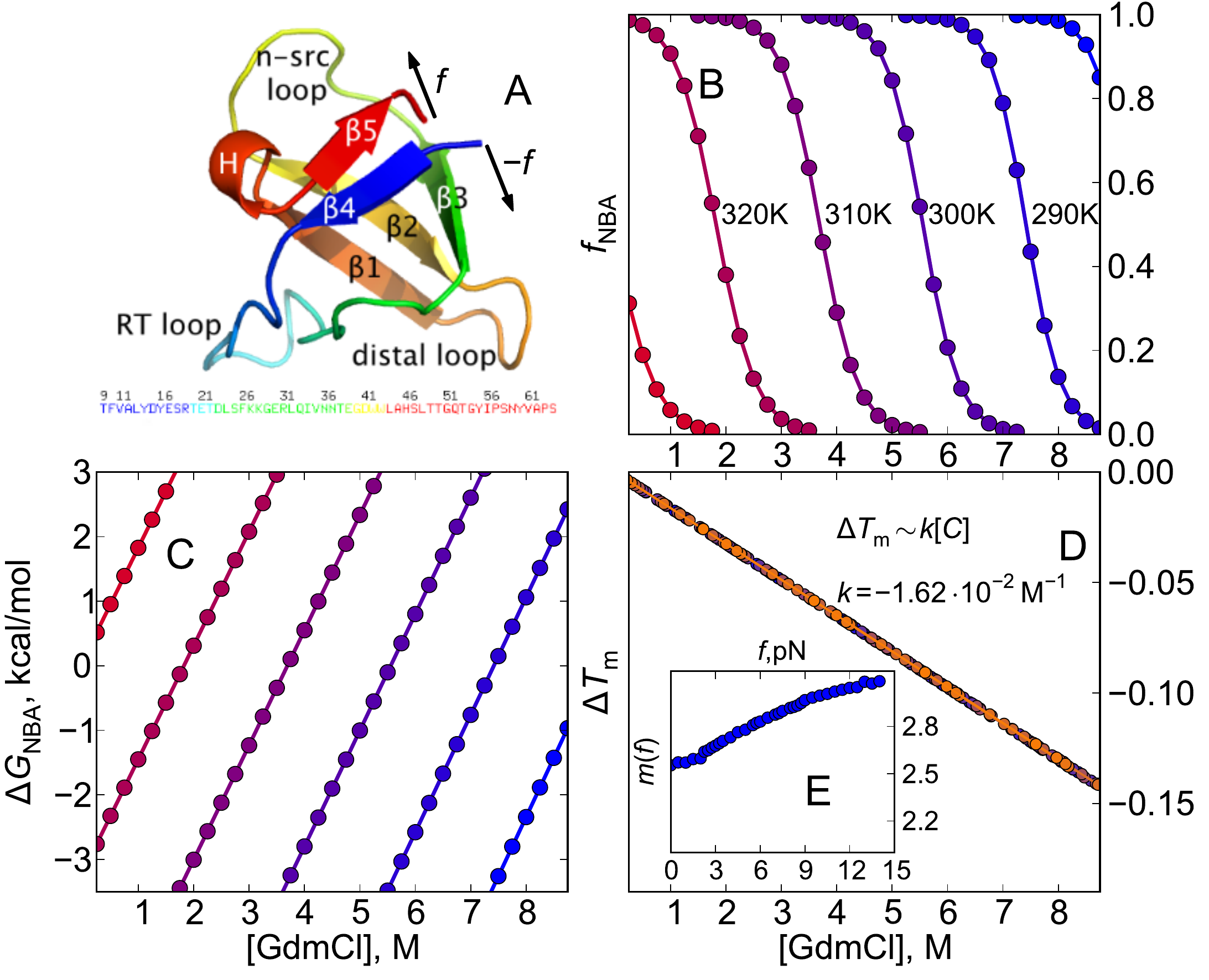}
       \caption{ Force-dependent thermodynamics as a function of temperature and denaturant concentration: (A) Crystal structure and sequence of the src SH3 domain (1SRL). (B) Guanidinium chloride titration curves at $f=7$ pN. (C) Protein stabilities as functions of [GdmCl] in the two-state picture, $\Delta G_\mathrm{NBA}=-k_BT\ln \frac{f_\mathrm{NBA}}{(1-f_\mathrm{NBA})}$. (D) Reduced melting temperatures $\Delta T_m=(T_m([C])-T_m([0]))/T_m([0])$ at different forces from 0 pN (blue) to 14 pN (orange). The data for all forces fall on a single line as explained in the main text; $[C]$ denotes the concentration of GdmCl, $\Delta T_m$ does not depend on $f$. (E) (inset) $m(f)$ (plotted in kcal/mol/M at $T=310$ K) depends weakly on $f$, with all of the dependence coming from the unfolded ensemble. 
\label{fig:titcurves}}
\end{figure}

\paragraph{SH3 stability as a function of perturbations}
The native structure of SH3 domains consists of five anti-parallel $\beta$-strands packed to form two perpendicular $\beta$-sheets. The strands are connected by the RT-loop, n-src loop, and a short distal loop (Fig.\ref{fig:titcurves}A). The N-terminal $\beta$-strand ($\beta$4 in Fig.\ref{fig:titcurves}A) participates in the hydrophobic core along with the strands $\beta$1, $\beta$2 and $\beta$3. In order to quantify the equilibrium response of SH3 to mechanical force ($f$), we performed multiple replica exchange low friction Langevin dynamics simulations (see Methods) by applying a fixed constant $f$ to the ends of the protein with and without denaturant.

To discriminate between the folded and unfolded states, that are populated in the equilibrium simulation trajectories, we use the order parameter $\chi$ (structural overlap function)
\begin{equation}
	\chi(\{\vect{r}\})=\frac{1}{M}\sum_{(i,j)}\Theta\left(\left||\vect{r_i}-\vect{r_j}|-|\vect{r_i^0}-\vect{r_j^0}|\right|-\Delta\right),
\end{equation}
where the sum is over the native contacts (as pairs of beads $(i,j)$), $M$ is the number of native contacts, $\Theta(x)$ is the Heaviside function, $\Delta=2\mathring{\mathrm{A}}$ is the tolerance in the definition of a contact, and $\vect{r_{i,j}}$ and $\vect{r_{i,j}^0}$, respectively, are the coordinates of the beads in a given conformation $\{\vect{r}\}$ and the native state. The histogram of $\chi$ exhibits a bimodal behavior, implying that SH3 folds in an apparent two state manner (see Fig.S1). The value of $\chi=\chi_c$ separating the native basin of attraction (NBA) from the unfolded ensemble, is obtained from analyzing the thermodynamics near the transition point, giving $\chi_c$=0.65. The fraction of proteins in the NBA $f_\mathrm{NBA}=\langle \Theta(\chi_c-\chi) \rangle,$ with the averaging being over the Hamiltonian, which includes the transfer energy contribution due to the presence of denaturant(see Methods).

The titration curves, plotting $f_\mathrm{NBA}$ as a function of $[C]$ (Fig.\ref{fig:titcurves}B) at different temperatures with $f=7$ pN, show that the midpoint of the transition, $C_m$, calculated using $f_\mathrm{NBA}([C_m];f)=0.5$, decreases sharply as $T$ increases. Interestingly, the results in Fig.\ref{fig:titcurves}B show that at $f\ne 0$, SH3 globally folds and unfolds reversibly in an apparent two state manner, just as in ensemble experiments at a fixed $T$\cite{FernandezEscamilla:2004ji,Grantcharova:1997ft,Grantcharova:2000wu,Riddle:1999bp,Martinez:1999iy}. Using the results in Fig.\ref{fig:titcurves}B, the computed $\Delta G_\mathrm{NBA}([C],T,f)=-k_\mathrm{B}T\ln\left[ f_\mathrm{NBA}/(1-f_\mathrm{NBA})\right]$ at a fixed $T$ with $f=7$ pN is shown in Fig.\ref{fig:titcurves}C. At a fixed $T$, the dependence on $[C]$ is given by $\Delta G_\mathrm{NBA}([C],T,f)=\Delta G_\mathrm{NBA}([0],T,f)+m(f)[C]$. Surprisingly, $m(f)$ is only weakly dependent on force because mechanical force does not perturb the intrinsic forces that determine the stability of proteins. Indeed, to the first approximation, $m$ should be proportional to the solvent accessible surface area, or $m \sim R_g^{2/3}$, and $R_g$ does not depend on $f$ while the protein is folded. From this perspective, use of $f$ is a natural way to perturb the protein as opposed to $T$ and $[C]$, which invariably alter the interactions involving proteins, the solvent, and denaturants. All of the $m(f)$ arises  from the increase in the solvent accessible surface ares of the unfolded ensemble (see Fig.S2), which grows as $[C]$ approaches $C_m$ from below. Linear response to $[C]$ is also indicated in the dependence of $T_{m}([C])=T_{m}([0],f)-\gamma(f) [C]$ at various $f$ values (Fig.\ref{fig:titcurves}D). By the reasoning given above we expect the reduced temperature $\Delta T_m=(T_{m}([C])-T_{m}([0]))/T_{m}([0])=k[C]$ to be proportional to $[C]$  at all the forces, with $k$ being independent of $f$. The melting temperature without the denaturant $T_{m}([0],f)$ depends on $f$, because the force changes the protein stability, but since it does not alter the interactions between the residues themselves, $k$ should not depend on $f$. The thick solid line in Fig.\ref{fig:titcurves}D, showing superposition of results at different $f$ values, indeed satisfies the expected ``scaling'' behavior with $k=-0.016\,\mathrm{M}^{-1}$ for all the $f$ values.

\begin{figure}[tbh]
	\centering
		\includegraphics[width=8.7cm]{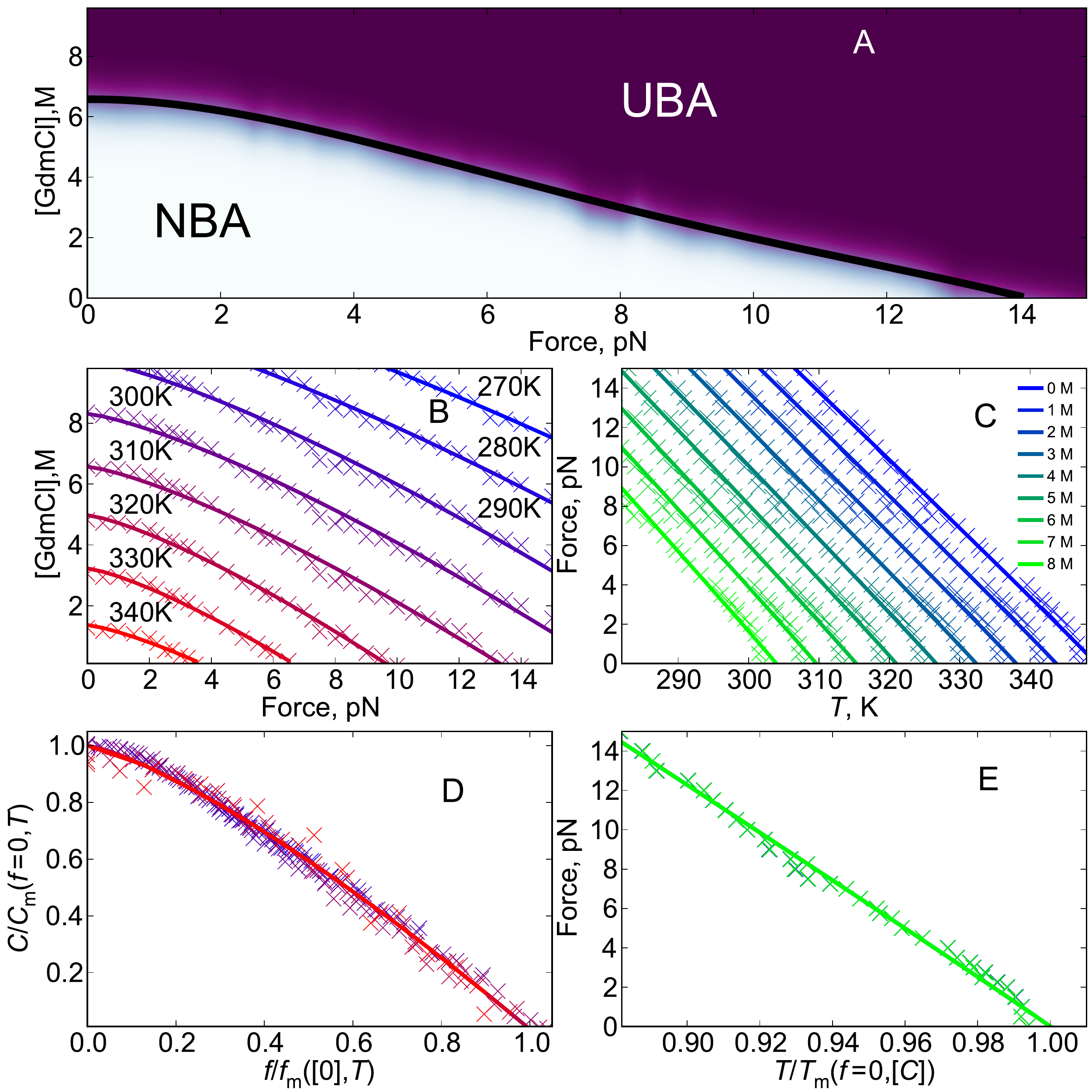}
       \caption{Phase diagram of the src SH3 domain: (A) Force-denaturant phase diagram of the src SH3 domain at T=310K. A fraction of conformations inside the native basin of attraction is color-coded from white ($f_\mathrm{NBA}$=1) to purple ($f_\mathrm{NBA}$=0). (B) Boundaries between the folded and unfolded regions at different temperatures. (C) Boundaries between the folded and unfolded regions at different GdmCl concentrations. (D) $(f,C)$ diagrams collapsed onto a master curve $C/C_0(T) = 1-(f/f_0(T))^\beta$, $\beta=1.25 \pm 0.1$. (E) $(f,T)$ diagrams collapsed onto a master curve $f/f_0([C]) = 1-(T/T_0([C]))^\alpha$.
\label{fig:pds}}
\end{figure} 

\paragraph{Phase diagrams}
From the titration curves at multiple forces and a fixed temperature ($T=310$K) we constructed the force-denaturant ($f,[C]$) phase diagram. Fig.\ref{fig:pds}A shows that the critical (melting) force $f_m([C])$ needed to unfold SH3 increases as $[C]$ decreases, reflecting the $[C]$-dependent stability of the native state. The boundary separating the NBA and unfolded basin of attraction(UBA) is relatively sharp, implying that SH3 behaves as a two-state folder in the $(f,[C])$ plane. The phase boundaries at other temperatures are shown in Fig.\ref{fig:pds}B. Similarly, the phase diagram as a function of $(f,T)$ calculated at different $[C]$ (see Fig.\ref{fig:pds}C) also shows a two-state behavior. Not surprisingly, the melting temperature and the critical (melting) force decreases as $[C]$ increases (Fig.\ref{fig:pds}B). For a fixed $T$, the phase boundary can be quantified using,

\begin{equation}
	C_m(f) \sim  C_0\left(1-\left(\frac{f}{f_0}\right)^\beta\right)
\label{eq:fcofT}
\end{equation}
where $C_m(f,T)$ is melting concentration, and $f_0(T)$ is the melting force at $[C]=0$. The parameters $f_0(T)$, $C_0(T)$ and $\beta$ should depend on the protein. From fitting we find $\beta=1.25 \pm 0.1$, estimating error using jackknife\cite{Efron:1982uw}. A similar fit ($f_m/f_0([C]) = 1-(T/T_m)^\alpha$) can be used to determine the family of curves at various values of $[C]$\cite{Klimov:1999up} (Figs.\ref{fig:pds}C and \ref{fig:pds}E). The solid lines in Figs.\ref{fig:pds}D and \ref{fig:pds}E show that all curves (given in Figs.\ref{fig:pds}B and \ref{fig:pds}C) collapse onto a master plot using the scaling functions above. The collapsed phase boundary in the restricted range of $(f,T)$ values in Fig.\ref{fig:pds}E appears linear, but we expect the master curve to be non-linear ($\alpha$ different from unity), which will be observable if the range of temperature is increased. However, such low $T$ values may not be physically relevant. Nevertheless, the finding that the phase boundary for various values of $[C]$ and $f$ collapse onto a single curve is surprising, and is amenable to experimental scrutiny. 

The behavior of phase boundaries here is analogous to that found in superconductors, which exhibits a sharp phase boundary in the $(H,T)$ plane described by the identical power law\cite{1402-4896-2012-T151-014029}. The analogy to superconductor in a magnetic field is appropriate, because in both cases the transition is likely to be first order. As the magnetic field ($H$) in the superconductor, in our case, force enters linearly into the Hamiltonian, but is coupled to microscopic coordinates in a complicated fashion, leading to non-trivial response to $f$ here or $H$ in superconductor. Similarly, our Hamiltonian is a linear function of $[C]$, which is coupled to a very complicated function of coordinates.

\begin{figure}[tbh]
	\centering
		\includegraphics[width=8.7cm]{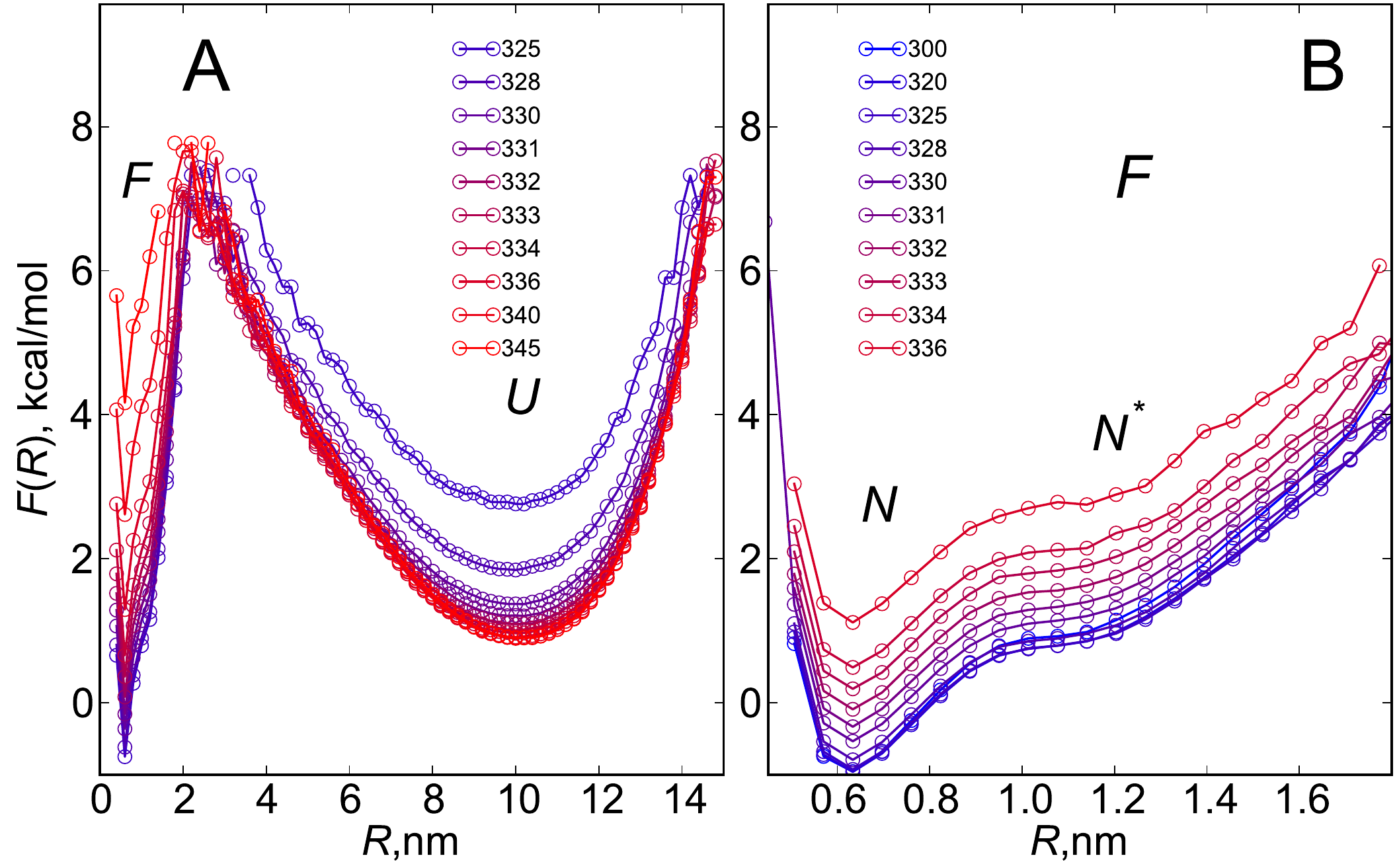}
       \caption{Free energy profiles at $f=7.5$ pN. (A) $F(R)$ at various temperatures, showing that globally, SH3 even under force unfolds in an apparent two-state manner. The transition state location $\Delta x^\ddagger_{N\rightarrow U} \approx 2$ nm is insensitive to temperature. (B) Fine structure in $F(R)$ shows the emergence of the $N^*$ state, which is prominent at high forces. Population of the $N^*$ state is between 3\% (at 300 K) and 6\% (at 340 K).  
\label{fig:fofree}}
\end{figure}

\begin{figure}[tbh]
	\centering								 
		\includegraphics[width=8.7cm]{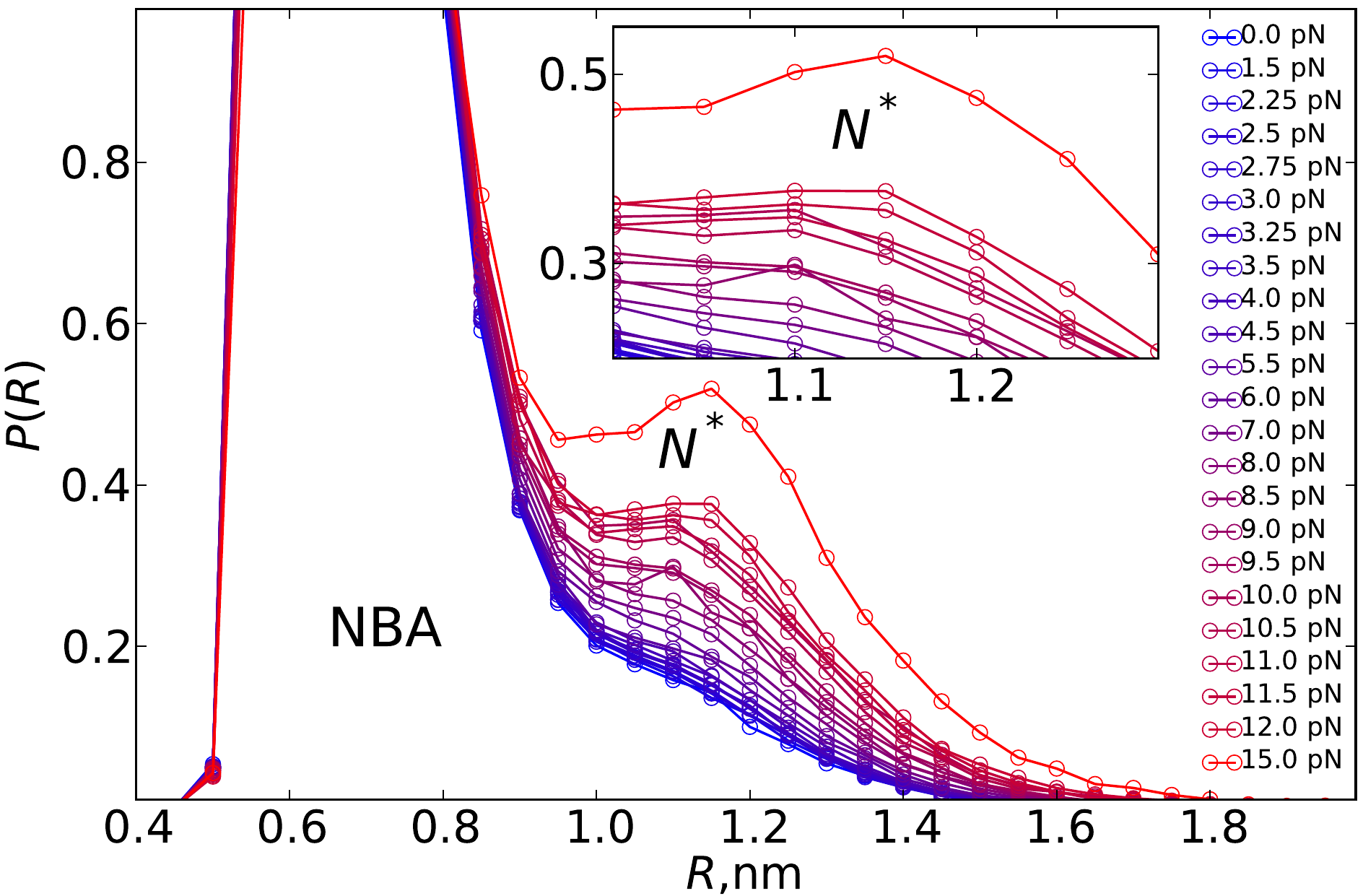}
       \caption{Distributions of the end-to-end distance, $R$, at different values of the stretching force at a fixed $T$=310K. The smaller peak corresponds to the aggregation-prone state $N^*$, which becomes more prominent as the magnitude of the force grows.
\label{fig:fshist}}
\end{figure}

\paragraph{N* state is visible in the presence of force}
 The global and experimentally accessible quantities in Fig.\ref{fig:titcurves} show that upon various perturbations ($T$, $[C]$ and $f$) SH3 folds in an apparent two state manner. A more nuanced picture emerges when the folding landscape is examined using a free energy profile $F(R)$, where $R$ is the extension of SH3 conjugate to $f$. Although not straightforward, $F(R)$ can be precisely calculated\cite{Hinczewski:2013kd} using folding trajectories generated in laser optical trap (LOT) experiments\cite{Greenleaf:2006bg,Stigler:2011ct}. The calculated free energy profile $F(R)=-k_BT\ln P(R)$, where $k_B$ is the Boltzmann constant, $P(R)$ is the distribution of $R$, exhibits a minimum at $R=1.15$ nm, which is distinct from the one at $R=0.68$ nm corresponding to the NBA (Figs.\ref{fig:fofree},\ref{fig:fshist}). The fine structure in $F(R)$ is subtle, and is not noticeable in other quantities such as the distribution of the radius of gyration (see Figs.S3-S6). The $R=1.15$ nm peak in $P(R)$ does not correspond to the globally unfolded state at $R \approx 10$ nm. Barely noticeable at $f=0$, the second minimum in $F(R)$ becomes increasingly pronounced as $f$ grows (Fig.\ref{fig:fshist}). In other words, mechanical force reveals an elusive state in the $F(R)$ profiles.

\paragraph{Structural characteristics of N*}
 In order to characterize the structure of the $N^*$ state corresponding to the peak at $R=1.15$ nm (Fig.\ref{fig:fshist}), we calculated the individual overlap parameters using $Q_{\beta 1,\beta 2}$, $Q_{\beta 2,\beta 3}$ and $Q_{\beta 4,\beta 5}$\cite{Hardin:2000ga} for each pair of $\beta$-strands that are part of a $\beta$-sheet using
\begin{equation}
Q(\{\vect{r}\})=1/M\sum \exp\left[-4(|\vect{r_i}-\vect{r_j}|-|\vect{r_i^0}-\vect{r_j^0}|)^2\right],
\end{equation} 
where the sum is over the contacts defining the $Q(\{\vect{r}\})$ (for instance, for $Q_{\beta 1,\beta 2}$ these would be pairs of beads, one from $\beta$1 and the other from $\beta$2 over all the beads in $\beta$1 and $\beta$2); $M$ is the number of such contacts, $\vect{r_{i,j}}$ are the coordinates of the beads in the conformation for which the $Q$ is calculated ($\{\vect{r}\}$), and  $\vect{r_{i,j}^0}$ are the corresponding coordinates in the native state.

\begin{figure}[tbh]
	\centering
		\includegraphics[width=8.7cm]{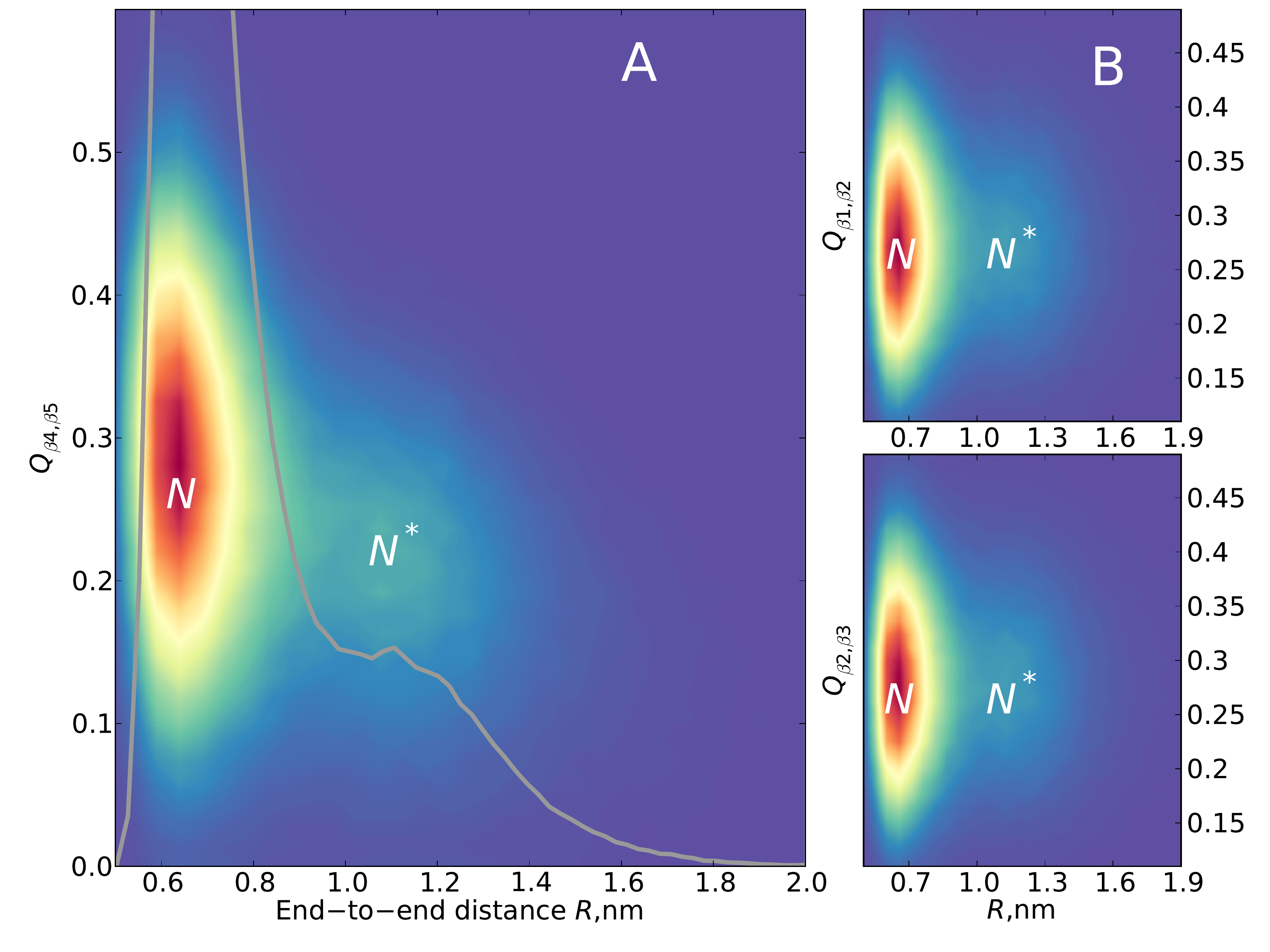}
       \caption{Structural characteristics of the $N^*$ state. Histograms of the conformations in the folded state at $f=10$ pN and $T=350 K$. The second, lower peak in $P(R)$ correlates with the melting of the $\beta 4$-$\beta 5$ sheet as shown by the fraction of contacts between the residues of those strands $Q_{\beta 4,\beta 5}$ (A). The y-axis on the 1D plot shows the probability density, while for the 2D plot it is $Q_{\beta 4,\beta 5}$, and the probability density is shown by color from blue (low) to red (high). $Q_{\beta 1,\beta 2}$ and $Q_{\beta 2,\beta 3}$ (B) remain the same in the $N^*$ state as in $N$.	
\label{fig:qb4b5reerub}}
\end{figure}

Two-dimensional histogram, in terms of the order parameters $(R,Q_{\beta 4,\beta 5})$ in Fig.\ref{fig:qb4b5reerub}A, shows that the second peak in the $P(R)$ profiles (Fig.\ref{fig:fshist}) corresponds to a smaller value of $Q_{\beta 4,\beta 5}$ compared to that found in the NBA. However, $Q_{\beta 1,\beta 2}$ and $Q_{\beta 2,\beta 3}$ retain their values in the native state (the higher peak) (Fig.\ref{fig:qb4b5reerub}B). Consequently, the second peak in $P(R)$ must reflect a state in which the $\beta$4-$\beta$5 sheet is melted (or disordered) with the rest of the native structure remaining intact. Because the $N^*$ structure is native-like in SH3, we surmise that it is a native substate, which is hidden at $f=0$. This assumption is corroborated by kinetic simulations showing that $N^*$ is frequently visited during the native dynamics. We identify the ensemble of conformations belonging to the second peak as the $N^*$ state that is aggregation prone (see below for comparison to experiments).

\begin{figure}[tbh]
	\centering
		\includegraphics[width=8.7cm]{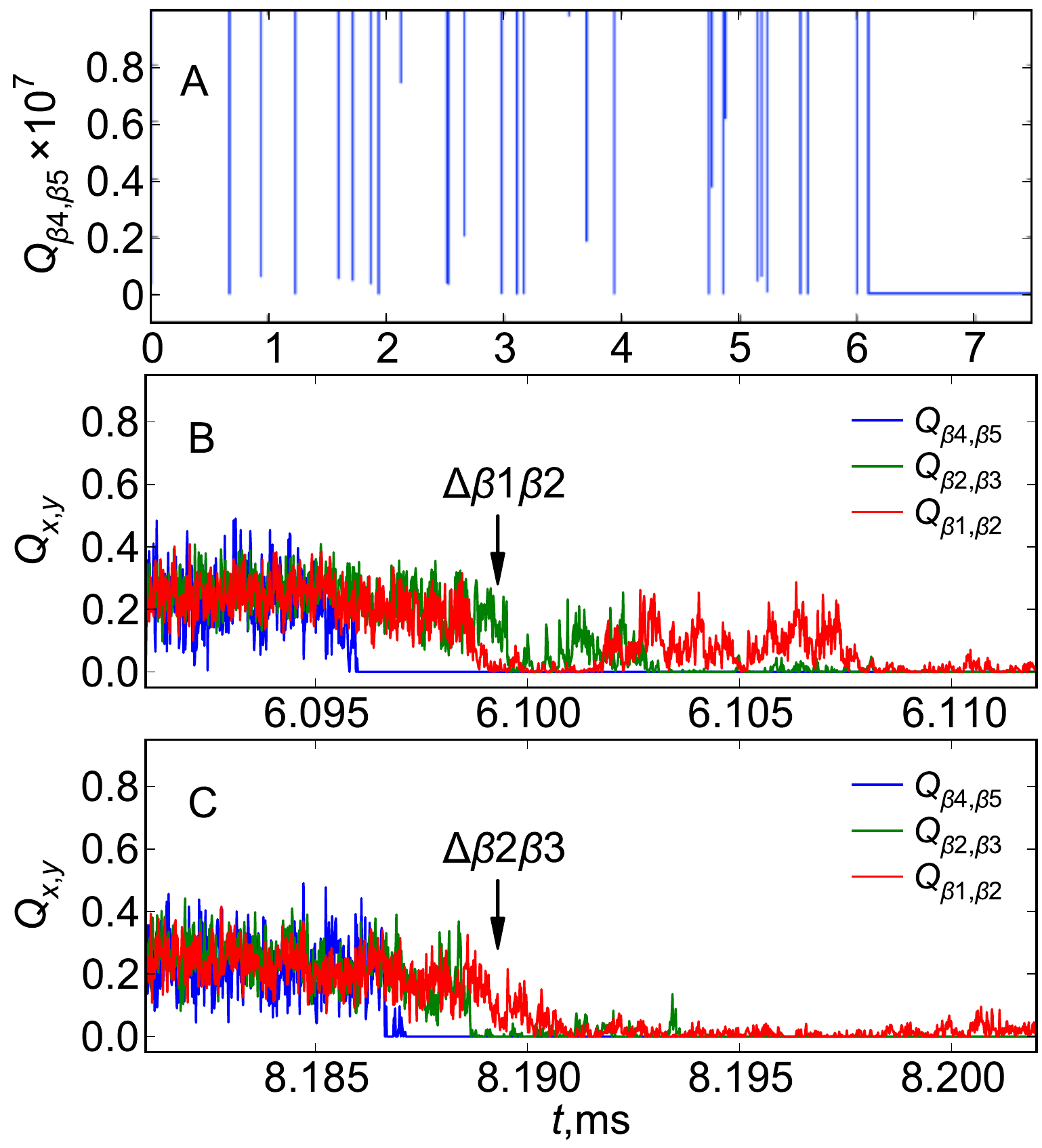}
       	\caption{Highlights of unfolding kinetics under stretching (trajectories at $f=10$ pN, $T=350$ K, $[C]=0$ are shown). (A) The $\beta$4-$\beta$5 sheet melts and reforms on the order of $10^{-5}$ s before completely unfolding. (B,C) When $\beta$4-$\beta$5 stays unfolded long enough so that contacts in the other $\beta$-sheet start breaking, the $\beta$1-2-3 sheet also melts completing the unfolding. Contacts between $\beta$1 and $\beta$2, and between $\beta$2 and $\beta$3 continue to transiently and partially reform even after reaching the unfolded state. $\beta$1-$\beta$2 breaks before the $\beta$2-$\beta$3 in trajectory shown in (B). The reverse sequence occurs in the trajectory shown in (C).
	\label{fig:kgraph}}
\end{figure}

\paragraph{Force-induced unfolding kinetics}
 In order to determine if remnants of the $N^*$ state arise in the force-unfolding kinetics of SH3 domain we performed Brownian dynamics simulations at high friction as described in Methods  at $T=340$, 350 and 360K, and $f=10$ pN. The observed unfolding times spanned a broad range of timescales (from hundreds of microseconds to over 10 milliseconds) (see Table S1 showing the unfolding times for individual trajectories). The sequence of events during unfolding was always the following. First the $\beta$4-$\beta$5 ruptures, populating the $N^*$ state. Following this event, in about 2-3$\mu$s, the $\beta 1$-$\beta 2$-$\beta 3$ sheet rips, taking a few more microseconds (Figs.\ref{fig:kgraph}B and Figs.\ref{fig:kgraph}C). In this process, contacts between either $\beta 2$-$\beta 3$, or $\beta 1$-$\beta 2$ can break first. Both $\beta 2$-$\beta 3$ and $\beta 1$-$\beta 2$ sheets continue to transiently and partially reform in the unfolded state. Thus, breaking of the $\beta 4$-$\beta 5$ sheet is not by itself rate limiting in the global unfolding of SH3 under these conditions. The sheet ruptures and reforms (Fig.\ref{fig:kgraph}) on the order of hundreds of microseconds (in agreement with the reported accessibility of the $N^*$ on the scale of milliseconds in the NMR experiment for Fyn SH3\cite{Neudecker:2012kn}), but only when it stays melted long enough for the other $\beta$ sheet to follow (about 2$\mu$s) does the protein globally unfold. The kinetic simulations show that $N^*$ is the same intermediate identified in equilibrium free energy profiles. Thus, in the folding landscape of SH3, the $N^*$ state is indeed a folding intermediate forming after the major folding barrier is crossed leading to formation of $N$ (formation of the $\beta$1-$\beta$2-$\beta$3 sheet). Since we observe unfolding rather folding, and $N^*$ is visited multiple times before global unfolding, it is natural to think of it as a native substate; $N^*$ is a native excitation around the folded state, and is visited during native dynamics even when the protein is thermodynamically in the folded state. After the initial event, there is a bifurcation in the unfolding pathways; in some of the trajectories $\beta 1$-$\beta 2$ ruptures first (Fig.\ref{fig:kgraph}B for example) while in others $\beta 2$-$\beta 3$ unfolds first (Fig.\ref{fig:kgraph}C). Thus, the kinetic trajectories also show a two-state behavior, with two possible unfolding pathways.

\begin{figure}
	\centering
		\includegraphics[width=8.7cm]{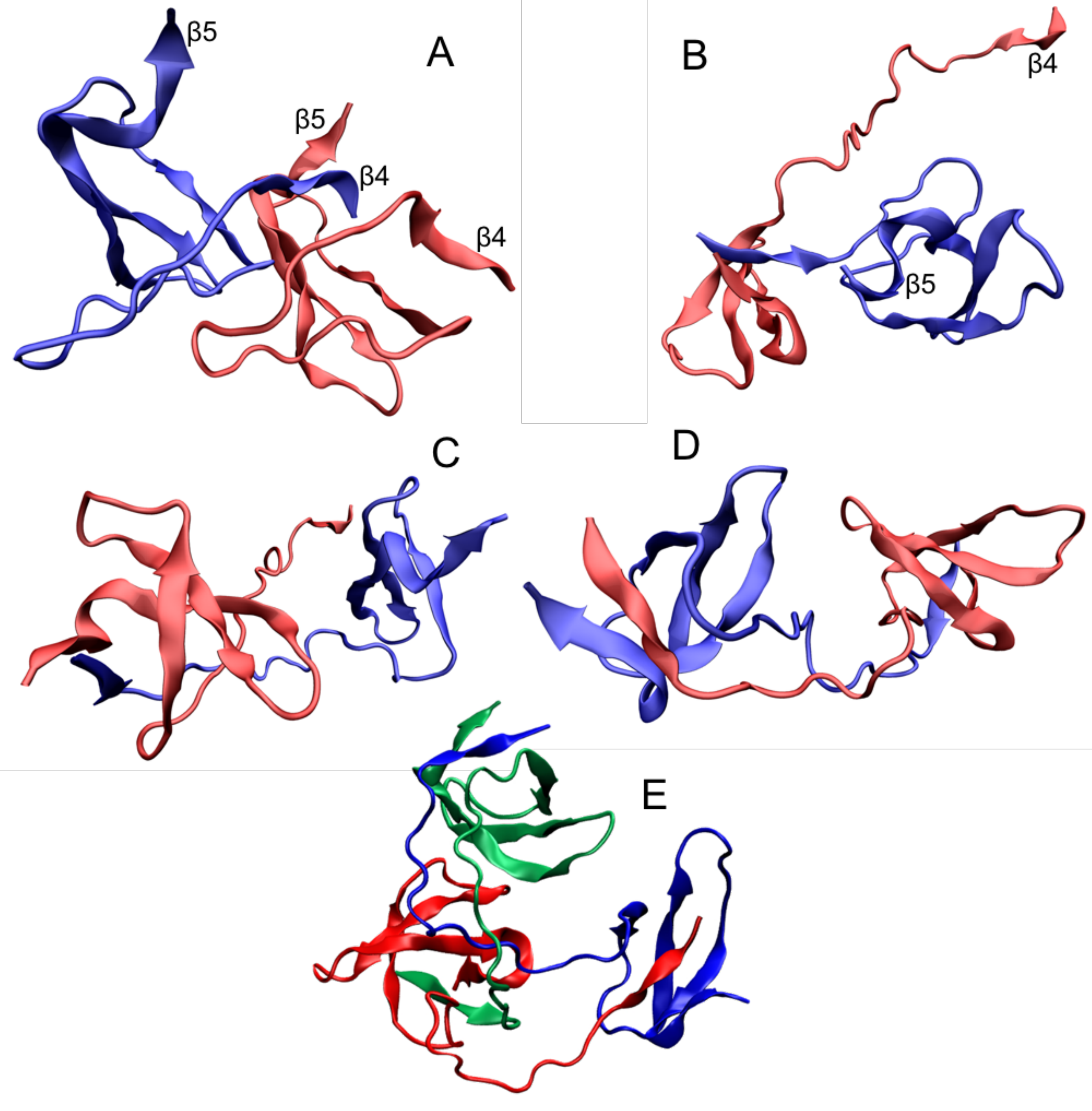}
       \caption{Stages of dimerization through domain swapping. (A) Starting configuration, corresponding to two monomers in $N^*$ conformation. (B) RT-loop of the red molecule unfolded with sticky $\beta 4$ on its end; the blue RT-loop starts to unfold. (C) RT-loops unfolded. Red $\beta 4$ continues diffusional motion around the blue monomer. (D) Red $\beta 4$ finds forms a $\beta$-sheet by associating with the blue $\beta 5$ in a domain-swapped dimer. (E) Trimer of the src SH3 domain formed during simulations of three SH3 molecules.  
\label{fig:dimerization}}
\end{figure}

\paragraph{Dimerization}
Our main hypothesis is that the excited $N^*$ state is prone to aggregation. In order to test this hypothesis, we have simulated the process of dimerization, starting from the two monomers in $N^*$ conformations, which were generated in the monomer simulations with one pair of $\beta 4$ and $\beta 5$ in contact, and the other pair on the opposite sides (Fig.\ref{fig:dimerization}A). We performed overdamped Brownian dynamics simulations at $T=277$K, controlling the concentration of the protein by constraining the distance between the centers of mass to 15 $\AA$ ($\sim R_g$ of a monomer)\cite{Dima:2002dj}.  On a relatively short time scale,  we observed dimerization, through the stages presented in simulation snapshots in Fig.\ref{fig:dimerization}B-D: partial unfolding of RT-loops, and the remaining $\beta 4$ finding the $\beta 5$ of the other molecule, forming the domain-swapped dimer (see \url{http://youtu.be/4a83Kv0t04c}). Thus, the most probable route to aggregation in this protein is through domain-swap mechanism, which was already established in a previous study\cite{Ding:2002tc}. Here, we explicitly show that domain swapped structures form readily by accessing a high energy $N^*$ state.

To investigate  whether this process of swapping the $\beta 4$ can provide a pathway to aggregation of multiple molecules, we also simulated addition of a third monomer to the dimer. We started from a configuration where dimer is not formed with one of the sticky $\beta 4$ strands being unstructured (Fig.\ref{fig:dimerization}C). With the same conditions for temperature and concentration as for dimer formation, we observed formation of a trimer (Fig.\ref{fig:dimerization}E and \url{http://youtu.be/2sNnQB0qfww}).

\section{Discussion} 
\paragraph{N* state as precursor to oligomerization}
 Theoretical arguments along with molecular dynamics simulations of peptides and proteins\cite{Thirumalai:2003wn,DeSimone:2012ij,Krobath12JMB,Rosenman:2013kk} have shown that aggregation is initiated when the protein (at least transiently) populates a high free energy $N^*$ state. In the last few years there have been several experiments on  proteins of different lengths, with no sequence or structural similarity, in which the $N^*$ states have been experimentally characterized\cite{Eichner:2009vq,DeSimone:2011gt,Neudecker:2012kn}. The closest example that is very similar to the $N^*$ state studied here arises in \textit{G.Gallus} Fyn SH3 domain\cite{Neudecker:2012kn}. Remarkably, Kay and coworkers identified using relaxation dispersion NMR experiments, a state with low population in A39V/N53P/V55L mutant Fyn SH3 under native conditions\cite{Neudecker:2012kn}, that is identical to our finding for the structurally similar src SH3 domain. The backbone structure of the folding intermediate determined for Fyn SH3 is similar to the native state everywhere except in regions adjacent to the N- and C- termini. A detailed comparison is provided in Fig.S7. Of particular note is that the C-terminal $\beta 5$ strand is disordered in the Fyn SH3 intermediate. The melting of the $\beta$4-$\beta$5 sheet leaves the N-terminal region exposed, which can trigger oligomer formation. They further corroborated the aggregation propensity of the region by preparing a truncation mutant, which is a mimic of the intermediate (or the $N^*$ state), and established that the resulting mutant forms amyloid-like fibrils with high $\beta$-strand content.

Interpreting the results of the NMR experiments on Fyn SH3 mutant using our findings, we conclude, that the folding intermediate of Fyn SH3 and the native substate identified in the src SH3 from the tyrosine kinase are the $N^*$ states. In other words, they are the ensembles of conformations in the spectrum of native excitations, that each monomer must populate in the course of oligomerization and fibril formation. Although not linked to any known disease, SH3 domains are known to aggregate in \textit{in vitro} experiments. In many proteins, containing the SH3 domain, for instance, the growth factor receptor-bound protein 2\cite{Lowenstein:1992vg} or phosphatidylinositol 3'-kinase\cite{Chen19941755}, SH3 is located at the N-terminus of the protein. It is likely that the mechanism of aggregation by exposure of the sticky N-terminal $\beta$-strand of SH3 might remain operational \textit{in vivo} as well.

\paragraph{$\beta_2$-microglobulin and acylphosphatase populate the N* under native conditions}
 Two other examples are also worth pointing out. In one of the earliest experiments, Radford and coworkers showed that aggregation of $\beta_2$-microglobulin into amyloid-like fibrils  occurs from a native-like ($N^*$) intermediate that has low population at equilibrium\cite{Jahn:2006dn,Eichner:2009vq}. Despite the large barrier created by the proline frozen in the \textit{cis} conformation, it was suggested that the $N^*$ state is part of the native state ensemble separated from the NBA by a large enough free energy gap that its population is low under native conditions. Finally, using NMR experiments and molecular dynamics with H/D exchange data as restraints, a free energy profile was reconstructed using RMSD of the $C_\alpha$ as the reaction coordinate for acylphosphatase was generated in the absence and presence of trifluoroethanol (TFE). These results suggest that a native-like $N^*$ state becomes visible at non-zero concentration of TFE. Just as in SH3 domains, interactions between two critical $\beta$-strands in acylphosphatase are disrupted exposing them to the solvent, thus making them aggregation prone\cite{Bemporad:2009ev}. Taken together, these results not only show that aggregation by populating $N^*$ state is a generic mechanism, but also reinforces earlier prediction that sequences of most natural proteins with predominantly $\beta$-sheet secondary structure have evolved so that intermolecular interactions between edge strands ($\beta 5$ in SH3 for example) are unfavorable\cite{Richardson:2002gv,Jahn:2006dn}.

\paragraph{Population of N* and fibril formation rates}
 In our previous work \cite{Li:2010tv} we showed that the time scale of fibril formation is related to $P_{N^*}$ (expressed in terms of percentage), the probability of populating the $N^*$ state as
\begin{equation}
	\tau_\mathrm{fib} \approx \lambda \tau_F \exp(-CP_{N^*}(T)),
\label{eq:aggtime}	
\end{equation} 
where $\tau_F$ is the folding time, $\lambda \approx 10^8$ and $C \approx 1$ \cite{Li:2010tv}. It is tempting to estimate $\tau_\mathrm{fib}$ for SH3 domain using the experimentally measured $\tau_F \approx 1.7 \cdot 10^{-2}$ s. The value of  $P_{N^*}(T)=2$\% for Fyn SH3 domain\cite{Neudecker:2012kn}, which coincides with the calculated value for src SH3 domain (see Figs. S8 and S9). Using $\tau_F$ and the estimate for $P_{N^*}(T)$  the above equation yields $\tau_\mathrm{fib}$ is on the order of 2-3 days, which is consistent with the measurements reported for Fyn SH3\cite{Neudecker:2012kn}.

A few implications for amyloidogenesis follows from the importance of $P_{N^*}$. (i) The fibril formation time scale should decrease dramatically as $P_{N^*}$ increases. (ii) Since $P_{N^*}$ (at a fixed temperature) depends on the free energy gap between $N^*$ and $N$ states, it follows that only those proteins for which there is a reasonable probability of being the $N^*$ state under physiological conditions would aggregate in relevant time scales. (iii) The temperature dependence of $\tau_\mathrm{fib}$ (Eq.\ref{eq:aggtime}) is highly non-trivial because $P_{N^*}(T)$ would be a maximum at some optimal temperature and is expected to decrease at low temperatures. (iv) It is the free energy gap between $N^*$ and the NBA, rather than the global stability, that determines the propensity of a protein to aggregate.

In our simulations force was applied to the ends of the protein. However, in simulations that we performed with applying force to the residues 9 (N-terminal end) and 59 (before the $\beta 5$ sheet in sequence), as in reported SMFS experiments with SH3\cite{Jagannathan:uw}, the $N^*$ is also accentuated by force, showing some robustness of the SMFS as a tool for highlighting the elusive states. We also note, that although not easily detectable, $N^*$ is present as the excitation in the native ensemble even at $f=0$, so many of the implications for fibril formation do not depend on force (including the pulling points). Thus, the detectability of $N^*$ to some extent does not depend on the pulling direction.

\begin{figure}
	\centering
		\includegraphics[width=8.7cm]{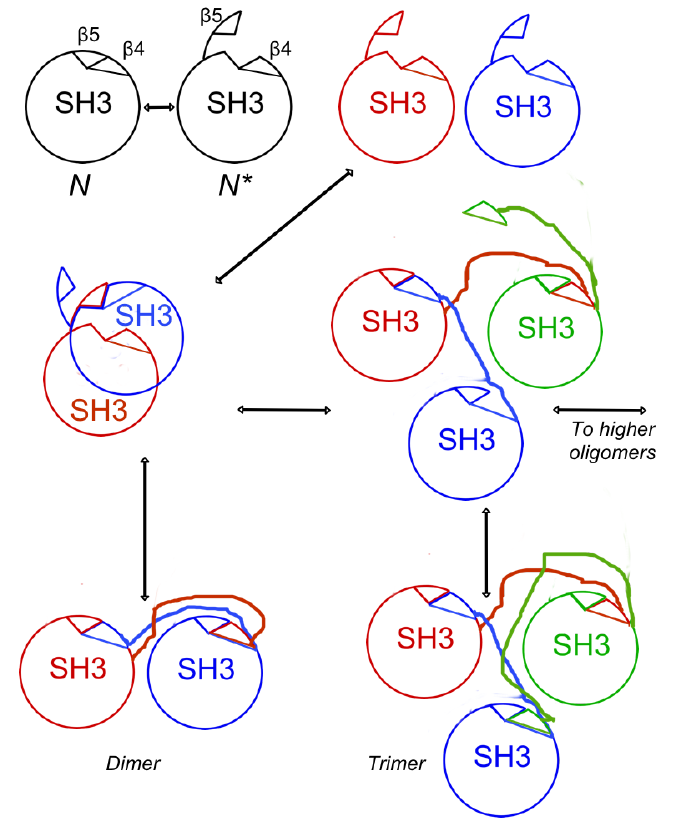}
       \caption{The pathway to aggregation through the $N^*$ state. First, if two molecules in $N^*$ state come close together, and one of the sticky $\beta 4$ makes contacts with the other molecule, the RT-loops will unfold. Eventually, the other $\beta 4$ can come into contact with $\beta 5$ of another molecule forming a domain-swapped dimer. This process can be repeated resulting in fibril formation.
\label{fig:aggregationpath}}
\end{figure}

\paragraph{From $N^*$ to oligomers to fibrils}
Our simulations of the process of oligomer formation starting from $N^*$ structures, support the finding that $N^*$ is aggregation prone. The ease of assembly of dimer and trimer by domain-swap mechanism allows for a route to aggregation in SH3 domain.  Our trimer simulation shows that when the dimer is not fully formed, the sticky $\beta 4$ dangling on the unfolded RT-loop, associated not with counterpart in the dimer but with a third molecule in $N^*$ state, leading to the formation of a  a trimer. In the formed dimer (or trimer), the contacts between $\beta 4$ and $\beta 5$  break once in a while, allowing for the sticky $\beta 4$ strand of another molecule to come into proximity, leading to the growth of higher order structures. This  domain-swapping (or rather $\beta$-strand swapping) mechanism can include an arbitrary number of molecules: $\beta 4_A \beta 5_A$ monomer, $\beta 4_A \beta 5_B + \beta 4_B \beta 5_A$ dimer, $\beta 4_A \beta 5_B + \beta 4_B \beta 5_C + \beta 4_C \beta 5_A$ trimer, $\beta 4_A \beta 5_B + \beta 4_B \beta 5_C + \beta 4_C \beta 5_D  + \beta 4_D \beta 5_A $ tetramer etc. (Fig.\ref{fig:aggregationpath}). The strand exchange mechanism shows that as the molecular weight of higher order structures increases, the higher is the probability that the $\beta 4$ strand at the end of the assembly will dangle on the RT-loop rather than form contacts with $\beta 5$, because  it becomes harder to accommodate the RT-loops within the oligomer of growing size.

The domain-swapping mechanism of src SH3 aggregation has been reported more than a decade ago\cite{Ding:2002tc}. Our main message here is the importance of native fluctuations of a monomer leading to the population of the aggregation-prone state $N^*$ and how this state can be gleaned by application of mechanical force. Our oligomerization simulations serve merely as an illustration of the principle that force can be used to access the $N^*$ states. Subsequent growth of oligomers can occur by domain swap mechanism as in SH3 or by other complex scenarios.

\paragraph{Rationale for using the coarse-grained models}
In all computer simulations it is important, but seldom practiced, to ascertain the extent to which the models capture reality. Our view is that the efficacy of the model can only be assessed by the ability to provide new insights and predict the outcomes of experiments. We have indeed produced a number of novel testable predictions using coarse-grained models. This level of falsifiable predictions cannot be currently achieved using all-atom simulations. We have shown in a number of studies that coarse-grained model used here has been remarkably successful in reproducing and predicting the force-stretching single-molecule experimental data in proteins\cite{Mickler:2007ky,Hyeon:2006gs,Hinczewski:2013kd,Hyeon:2011is} and RNA\cite{Lin:2008p12431,Hyeon:2006gs} including predictions that have been later validated quantitatively. The SOP-SC model used in this work is a more refined version of SOP. It has reproduced experimental results for the denaturant induced unfolding\cite{Liu:2011jc,Reddy:2012dg}. The choice of the SOP-SC model, therefore, is very adequate for studying the questions we have answered in this work.

\section{Conclusions}
We have established, using \textit{in silico} force spectroscopy of the src SH3 domain, that response of native proteins to mechanical force  provides quantitative insights into protein aggregation. The force-denaturant phase diagram for the SH3 domain shows that the boundary separating the folded and unfolded states is universal (Eq.\ref{eq:fcofT}) implying that determination of $C_0$, $f_0$ and $\beta$ is sufficient to predict the phase diagram at any temperature. 

Of particular importance is the demonstration that mechanical force can be used to elucidate a fine structure in the organization of the native state. In particular, force highlights the aggregation-prone $N^*$ state in the native ensemble of SH3, by increasing its stability. The structure of the $N^*$ state in src SH3 domain is identical to an aggregation-prone folding intermediate recently found in the Fyn SH3 domain in a relaxation dispersion NMR experiment\cite{Neudecker:2012kn}. Simulations of kinetics of unfolding under  force shows that the $N^*$ state has to be formed and linger for a substantial time period prior to global unfolding. Our work shows that force spectroscopy in combination with simulations, carried out under the same conditions as experiments, is a viable technique to characterize important substates in the spectrum of native excitations even if they have low populations, as is the case in SH3 domain.

A byproduct of our study is that force (or more generally stress) can enhance the probability of oligomer and fibril formation. This implies that proteins that operate against load (motor proteins for example) have evolved to minimize the tendency to sample $N^*$ states by ensuring that the free energy gap separating the $N^*$ and the native state is large.

It follows from our work that aggregation propensity of folded proteins under native conditions is determined by the accessibility of the $N^*$ state, which in SH3 domains involves partial unfolding of the folded state. Thus, it is the free energy gap separating the NBA and $N^*$ state ($\Delta G_{\mathrm{NBA},N^*}$), rather than the global stability, which plays a critical role in protein aggregation. A corollary of this prediction is that mutations that increase (decrease) $\Delta G_{\mathrm{NBA},N^*}$ would decrease (increase) aggregation probability. Protein aggregation, therefore, depends not only on sequence, but also on $\Delta G_{\mathrm{NBA},N^*}$, which can be manipulated by changing external conditions.

Our work also shows that by knowing the structure of $N^*$ state it might be possible to construct structural models for oligomers, as we have done for the src SH3 domain. The generality of this approach requires further work.

\section{Methods}

\paragraph{Model}
  Our simulations were performed using the self-organized polymer model with side-chains (SOP-SC)\cite{Liu:2011jc} for the protein. In the SOP-SC model each amino-acid is represented using two interaction centers. The energy function in the SOP-SC representation of a polypeptide chain is taken to be,
\begin{equation}
	\mathcal{H}=U_\mathrm{LJ}^\mathrm{nat}+U_\mathrm{SS}^\mathrm{nnat}+U_\mathrm{SS}^\mathrm{neib}+U_\mathrm{FENE}.
	\label{eq:hamilt}
\end{equation}

The $U_\mathrm{LJ}^\mathrm{nat}$ potential is based on the native structure (PDB structure in our case), which assigns an attraction between the residues that are in contact in the native state and are at least 2 residues apart along the chain. The form of $U_\mathrm{LJ}^\mathrm{nat}$ is,
\begin{equation}
	\begin{split}
	U_\mathrm{LJ}^\mathrm{nat}=&\sum\limits_{(j-i)>2}\varepsilon_\mathrm{bb}f_\mathrm{LJ}(\vect{r}_i^\mathrm{bb},\vect{r}_j^\mathrm{bb})+\\
	+&\sum\limits_{(j-i)>2}\varepsilon_\mathrm{ss}|\epsilon_{ij}-0.7|f_\mathrm{LJ}(\vect{r}_i^\mathrm{ss},\vect{r}_j^\mathrm{ss})+\\
	+&\sum\limits_{|j-i|>2}\varepsilon_\mathrm{bs}f_\mathrm{LJ}(\vect{r}_i^\mathrm{bs},\vect{r}_j^\mathrm{bs}),
\end{split}
\end{equation}
where the sums are over the residues.  We use the Lennard-Jones function for the beads that are in contact in the native state (with the minimum at the native distance):
\begin{equation}
	f_\mathrm{LJ}(\vect{r}_1,\vect{r}_2)=\left[\left(\frac{r_{12}^0}{|\vect{r}_1-\vect{r_2}|}\right)^{12}-2\left(\frac{r_{12}^0}{|\vect{r}_1-\vect{r_2}|}\right)^6\right]\Delta(r^0_{12}),
\end{equation}
where $\Delta=1$ if the native distance in the PDB $r^0_{12}<R_c$ (we use $R_c=8\mathrm{\mathring{A}}$), and $\Delta=0$ otherwise. We use the Betancourt--Thirumalai statistical potential for  $\epsilon_{ij}$\cite{Betancourt:1999en}.  Superscripts $\mathrm{b}$ and $\mathrm{s}$ stand for backbone bead and side chain bead, respectively.

Non-native interactions are modeled by soft sphere interactions:
\begin{equation}
	\begin{split}
	U_\mathrm{SS}^\mathrm{nnat} = & \sum\limits_{(j-i)>2}\varepsilon_{l}\left(\frac{\sigma^\mathrm{bb}}{r_{ij}^\mathrm{bb}}\right)^6(1-\Delta(r_{ij}^{\mathrm{bb},0}))+\\
	+&\sum\limits_{(j-i)>2}\varepsilon_{l}\left(\frac{\sigma_{ij}^\mathrm{ss}}{r_{ij}^\mathrm{ss}}\right)^6(1-\Delta(r_{ij}^{\mathrm{ss},0}))+\\
	+&\sum\limits_{|j-i|>2}\varepsilon_{l}\left(\frac{\sigma_{ij}^\mathrm{bs}}{r_{ij}^\mathrm{bs}}\right)^6(1-\Delta(r_{ij}^{\mathrm{bs},0})),
	\end{split}
\end{equation}
where $\sigma^\mathrm{bb}=3.8\mathrm{\mathring{A}}$ is the average distance between neighboring $C_\alpha$ atoms, and $\sigma^{\alpha \beta}_{ij}$ is the sum of van der Waals radii of corresponding beads (backbone or side-chain depending on the amino acid).

Repulsive interactions between neighboring residues stabilize local secondary structure:
\begin{equation}
	U^\mathrm{neib}_{SS}=
	\sum\limits_{(j-i)\leq 2}
	\varepsilon_{l}\left(\frac{\sigma^\mathrm{bb}}{r_{ij}^\mathrm{bb}}\right)^6+\varepsilon_{l}\left(\frac{\sigma^\mathrm{ss}_{ij}}{r_{ij}^\mathrm{ss}}\right)^6+\sum\limits_{|j-i|\leq 2}
	\varepsilon_{l}\left(\frac{\sigma^\mathrm{bs}_{ij}}{r_{ij}^\mathrm{bs}}\right)^6.
\end{equation}
The values of all $\sigma$ and $\varepsilon$ are the same as used in \cite{Liu:2011jc}.

The chain connectivity is described by $U_\mathrm{FENE}$, or finite extensible nonlinear elastic potential:
\begin{equation}
	U_\mathrm{FENE}=\sum\limits_{\mathrm{bonds}}\frac{k}{2}R_0^2\log\left(1-\frac{l-l_0}{R_0^2}\right),
\end{equation}
where the summation is over all bonds (whether between two backbone beads or between a backbone and a sidechain bead), $l$ is the distance between the two bonded beads and $l_0$ is the length of the bond in the native state. We used spring constant $k=20\, \mathrm{kcal/mol/\mathring{A}^2}$ and $R_0=2\,\mathrm{\mathring{A}}$.

\paragraph{Model for denaturant}
 The molecular transfer model\cite{OBrien:2008p10988,Liu:2011jc,Liu:2012iw} is a phenomenological way to account for the effect of denaturants, where the effective interactions involving denaturant represent the free energy of transfer of a residue or backbone from pure water to the aqueous solution of the denaturant with concentration $[C]$. The transfer free energy term is a sum of terms proportional to the solvent accessible surface areas of individual residues (backbone and sidechain):
\begin{equation}
	\begin{split}
	\mathcal{H}_{[C]}(\vect{r})=&\mathcal{H}_{[0]}(\vect{r})+\Delta G(\vect{r},[C])=\\=&\mathcal{H}_{[0]}(\vect{r})+\sum\limits_k\delta g(k,[C])\frac{\alpha_k(\vect{r})}{\alpha_{\mathrm{Gly}-k-\mathrm{Gly}}},
	\end{split}
	\label{eq:mtmhamil}
\end{equation}
where the sum is over all backbone and sidechain interaction centers, $\delta g(k,[C])=m_k[C]+b_k$ is the transfer free energy of the bead $k$, and $m_k$ and $b_k$ are different for the backbone and sidechains, and depend on the residue identity. The solvent accesible surface area (SASA) is $\alpha_k(\vect{r})$ for bead $k$ in the protein conformation $\vect{r}$, and $\alpha_{\mathrm{Gly}-k-\mathrm{Gly}}$ is the SASA of the bead $k$ in the tripeptide Gly-$k$-Gly.  We used the same values of $m_k$, $b_k$ and $\alpha_{\mathrm{Gly}-k-\mathrm{Gly}}$ as in \cite{Liu:2011jc}. We used the procedure described in \cite{Hayryan:2005dg} to calculate the SASA.

Calculating SASA is computationally expensive. It is possible to avoid calculating $\Delta G(\vect{r},[C])$ at every simulation step, if only  equilibrium properties are of interest. If $\mathcal{H}_{[C]}$ is the full Hamiltonian, $\mathcal{H}_0$, and $-\Delta G(\vect{r},[C])$ is the restriction potential similar to the one used in umbrella sampling simulation, it is possible  to perform converged simulations using the Hamiltonian $\mathcal{H}_{[0]}$. The properties of the system under the unperturbed Hamiltonian $\mathcal{H}_0=\mathcal{H}_{[C]}$ can then be calculated using the weighted histogram analysis method (WHAM) equations\cite{WHAM}.

\paragraph{Multichain interactions}
The non-bonded interactions between beads on different chains were the same as if they were on the same chain. That is, if two residues, e.g. V61 and V11 form native contact, then V61 and V11 have the same native interaction (with $\varepsilon$s and $\sigma$s for valine-valine) even if they belong to different chains. Two residues that interact non-natively (i.e. via soft sphere repulsion) also behave the same way on different chains. Similar model for multichain interaction has been employed by Ding et al\cite{Ding:2002tc} for studying of aggregation of src SH3, that is the native interactions were the same between the residues whether on the same or different chains. The differences were that Ding et al. used pure G\-{o} model and added hydrogen bonds between the backbone atoms of different molecules.

\paragraph{Simulations}
 We used Large-scale Atomic/Molecular Massively Parallel Simulator (LAMMPS)  \cite{LAMMPS} to perform the simulations, enhancing it for the SOP-SC model. 
Langevin dynamics is performed as the numerical solution of the Langevin equation for every protein bead:
\begin{equation}
	m \vect{\dot v}=-\zeta \vect{v} + \vect{F_c} + \vect{F_r}(t),
	\label{eq:Langevin}
\end{equation}
where $m$ is the mass of the bead, $\vect{v}(t)=\vect{\dot r}$; $\vect{F_c}$ are conservative forces acting on the bead (interactions with the other bead), $\vect{F_c}=\partial\mathcal{H}/\partial{\vect{r}}$; $\vect{F_r}$ is the white noise random force with $\langle F_r^{i}(t) \rangle=0$ and  $\langle F_r^{i}(t)F_r^{j}(t^\prime) \rangle=2\zeta kT\delta_{ij}\delta(t-t^\prime)$. We integrated the Langevin equations using Verlet algorithm as implemented in LAMMPS\cite{PhysRev.159.98,LAMMPS}. The timescale in Eq.\ref{eq:Langevin} is defined by $\zeta/m$. The relevant timescale for low friction limit, where inertial term is significant in SOP-SC model is $\tau_L=\sqrt{m\sigma^2/\varepsilon_l}=2$ ps. For low friction limit (when simulating thermodynamics) we use $\zeta=0.05 m/\tau_L$.
We used LAMMPS facilities for the replica exchange simulations. We made random exchange attempts at 100$\tau_L$ intervals, with the same intervals between saving the trajectory snapshots for analysis (to calculate the transfer energies).
For the overdamped limit (when simulating kinetics), where the inertial term in the Langevin equation is negligible, the relevant timescale is $\tau_H=\zeta \sigma^2/kT$. For high friction we use $\zeta=50 m/\tau_L$, yielding $\tau_H \approx 150$ ps\cite{Veitshans:1997tz,Liu:2011jc}.

For calculating phase diagrams, we have performed replica exchange simulations at low friction and at the following fixed values of $f$: 0, 0.5,  1, 1.5, 2, 2.25, 2.5, 2.75, 3, 3.25, 3.5, 4, 4.5, 5, 5.5, 5.75, 6, 6.5, 7, 7.25, 7.5, 8, 8.25, 8.5, 9, 9.5, 10, 10.5, 11, 11.5, 12, 12.5, 13, 13.5, 14, 15 pN, at $[C]=0$ and used the sampled conformations at various temperatures to infer the distribution of NBA and denatured states at particular values of $C$ and $T$ for every $f$, using WHAM with the transfer energy in place of biasing potential as detailed in the subsection below. From individual replicas we calculated distributions and free energy profiles as a function of end-to-end distance $R$ for various values of $f$ (at fixed $T$ and $[C]=0$) and $T$ (for fixed $f$ and $[C]=0$)

For calculating unfolding kinetics, we ran 16 trajectories at 3 different temperatures, as detailed in the Table S1.

To illustrate dimerization and path to fibril formation we started the simulation with two molecules in $N^*$ conformation. We fixed the distance between their centers of mass at $\sim$ 15 $\AA$ ($\sim R_g$) with a parabolic potential (as a way to simulate high concentration) and ran Brownian dynamics at T=277K until observing the formation of a dimer, just after about 100 $\mu$s. The whole process is shown in the Supplementary Movie 1.

For the trimer formation, we started from a snapshot from the dimerization trajectory and added a third molecule in $N^*$ state. We constrained the three distances between three centers of mass to the same $\sim$ 15 $\AA$. We observed the formation of trimer (that is, the last $\beta 4$ finding its spot on another molecule next to its $\beta 5$) after about 1 ms. Supplementary Movie 2 shows $\sim 50 \mu$s preceding the formation of a trimer.

\paragraph{WHAM for equilibrium properties}
 For simulations with the Hamiltonian
\begin{equation}
	\mathcal{H}(\vect{r})=\mathcal{H}_0(\vect{r})+\lambda V_\mathrm{rest}(\vect{r}),
	\label{eq:whamham}
\end{equation}
where $V_\mathrm{rest}$ is the restriction potential with a coupling constant $\lambda$, WHAM can be used to calculate equilibrium properties of the system with Hamiltonian $\mathcal{H}_0(\vect{r})$. In case of MTM Eq.\ref{eq:whamham} becomes: 
\begin{equation}
	\mathcal{H}_{[0]}(\vect{r})=\mathcal{H}_{[C]}(\vect{r})-\Delta G(\vect{r},[C]).
\end{equation}
The unnormalized probability density of an observable $\xi$ from $M$ simulations performed using $\mathcal{H}_{[0]}$ with $i$-th simulation carried out at temperature $T_i$ is\cite{WHAM}:
\begin{equation}
	P(\xi,T_i,[C])=\frac{\sum\limits_{k=1}^M N_k(\xi)\exp\left({-\mathcal{H}_{[C]}/k_\mathrm{B}T_i}\right)}{\sum\limits_{m=1}^Mn_m\exp\left(f_m-\mathcal{H}_{[0]}/k_\mathrm{B}T_m\right)},
	\label{eq:whamprob}
\end{equation}
where $N_k(\xi)$ is the histogram of $\xi$ in $k$-th simulation, $n_m$ is the number of snapshots (conformations) analyzed for $m$-th simulation. The free energy of the $i$-th simulation, $f_i$ is:
\begin{equation}
	f_i=-\log\sum\limits_\xi P(\xi,T_i,[0]),
	\label{eq:whamf}
\end{equation}
which can be found self-consistently using:
\begin{equation}
	f_i=-\log\sum\limits_{k=1}^M\sum\limits_{t=1}^{n_k}\frac{\exp\left({-\mathcal{H}_{[0]}(k,t)/k_\mathrm{B}T_i}\right)}{\sum\limits_{m=1}^Mn_m\exp\left(f_m-\mathcal{H}_{[0]}(k,t)/k_\mathrm{B}T_m\right)}.
	\label{eq:freeeit}
\end{equation}
In Eq.\ref{eq:freeeit}, $\mathcal{H}_{[0]}(k,t)$ is the potential energy of the $t$-th snapshot from the $k$-th simulation (in the absence of denaturants).

After calculating the $f_i$ the averages for the MTM Hamiltionian (\ref{eq:mtmhamil}) can be obtained using (\ref{eq:whamprob}):

\begin{equation}
	\begin{split}
	\langle \xi([C_i],T_i)\rangle = & \\ =\frac{1}{Z([C_i],T_i)} \times & \sum\limits_{k=1}^M\sum\limits_{t=1}^{n_k}\frac{\xi(k,t)\exp\left({-\mathcal{H}_{[C_i]}(k,t)/k_\mathrm{B}T_i}\right)}{\sum\limits_{m=1}^Mn_m\exp\left(f_m-\mathcal{H}_{[0]}(k,t)/k_\mathrm{B}T_m\right)},
	\end{split}
\end{equation}
with the partition function
\begin{equation}
	Z([C_i],T_i)=\sum\limits_{k=1}^M\sum\limits_{t=1}^{n_k}\frac{\exp\left({-\mathcal{H}_{[C_i]}(k,t)/k_\mathrm{B}T_i}\right)}{\sum\limits_{m=1}^Mn_m\exp\left(f_m-\mathcal{H}_{[0]}(k,t)/k_\mathrm{B}T_m\right)},
\end{equation} 
where ${H}_{[C_i]}(k,t)={H}_{[0]}(k,t)+\Delta G(k,t,[C_i])$, and $\xi(k,t)$, $ {H}_{[0]}(k,t)$ and $\Delta G(k,t,[C_i])$ are the values of $\xi$, potential energy and transfer energy to the solution of concentration $[C_i]$ in the $t$-th snapshot of $k$-th simulation.


\makeatletter 
\def\tagform@#1{\maketag@@@{(S\ignorespaces#1\unskip\@@italiccorr)}}
\makeatother

\setcounter{figure}{0}

\makeatletter \renewcommand{\fnum@figure}
{\figurename~S\thefigure}
\makeatother

\makeatletter \renewcommand{\fnum@table}
{\tablename~S\thetable}
\makeatother

\onecolumn
\section{Supplementary Figures and Tables}

\begin{table}[h]
	\begin{center}
\caption{\label{t:times} Unfolding times for individual trajectories in kinetic simulations at $f=10\,\mathrm{pN}$, $[C]=0$ and various temperatures. Four trajectories were run at 340 K for 18 ms, which is too short to observe  unfolding of src SH3. The first passage unfolding times for eight trajectories at 350 K and four at 360 K  are listed. }

\begin{tabular}{c|c}
\hline
$T$			& $\tau_\mathrm{unf}$, ms \\
\hline
340 K&  \textgreater 18  (4 trajectories)\\
350 K&7.4,
9.0,
2.0,
6.5,
3.2,
6.7,
4.3,
8.2\\
360 K&5.3,
1.0,
0.71,
1.0
\\
\hline
\end{tabular}
\end{center}

\end{table}

\newpage

\begin{figure}
	\centering
		\includegraphics[width=8.7cm]{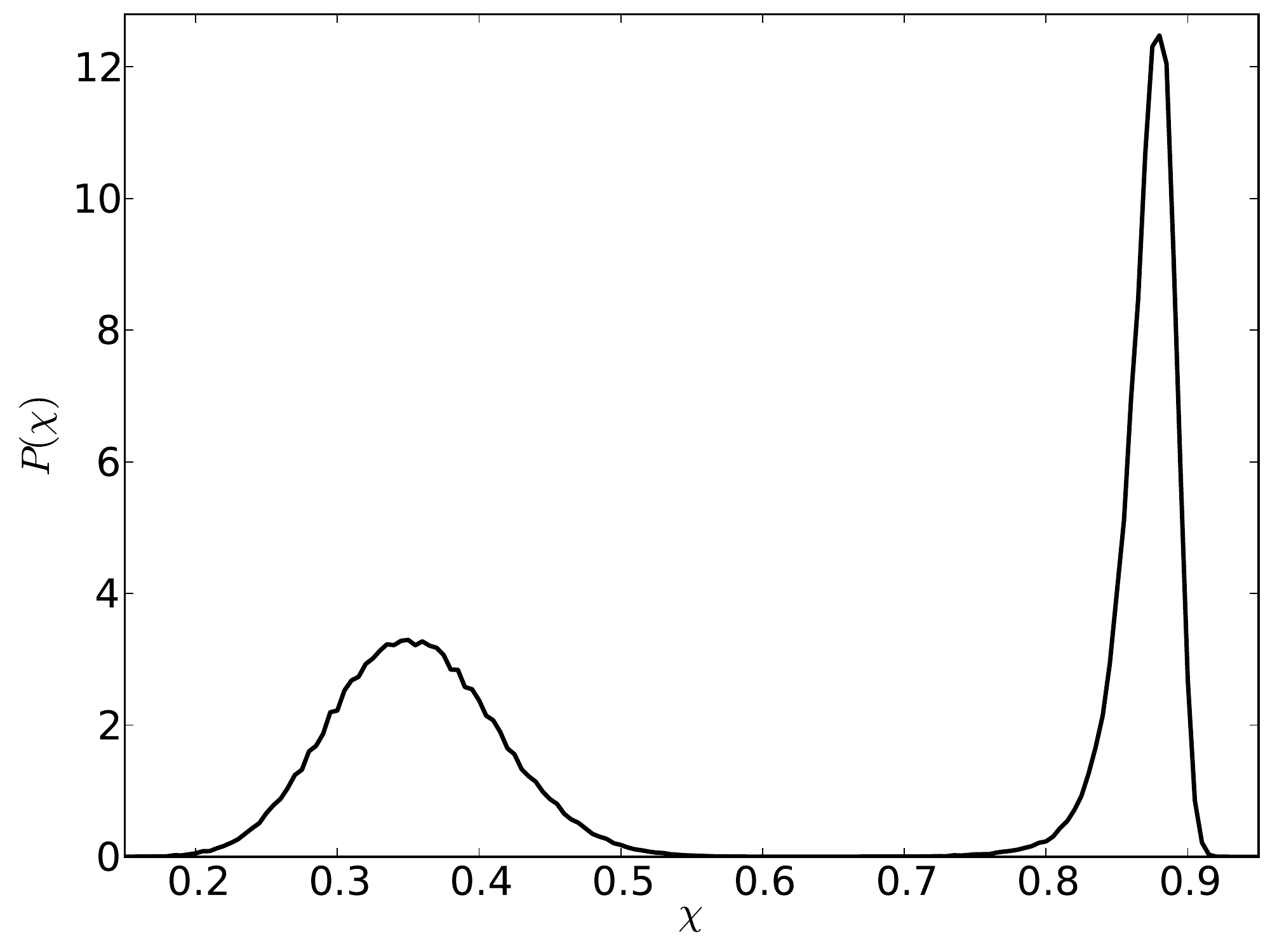}
       \caption{From analyzing thermodynamics at $f=7.5$ pN, we find that $T=331$ K is the melting temperature at this force value. Histogram of the overlap parameter $\chi$ at these parameter values (shown here) is bimodal, showing an apparent two-state behavior. At the transition point, $\int\limits_0^{\chi_c} P(\chi) d\chi=\int\limits_{\chi_c}^1 P(\chi) d\chi$.
\label{fig:chihistbimodal}}
\end{figure}

\begin{figure}
	\centering
		\includegraphics[width=8.7cm]{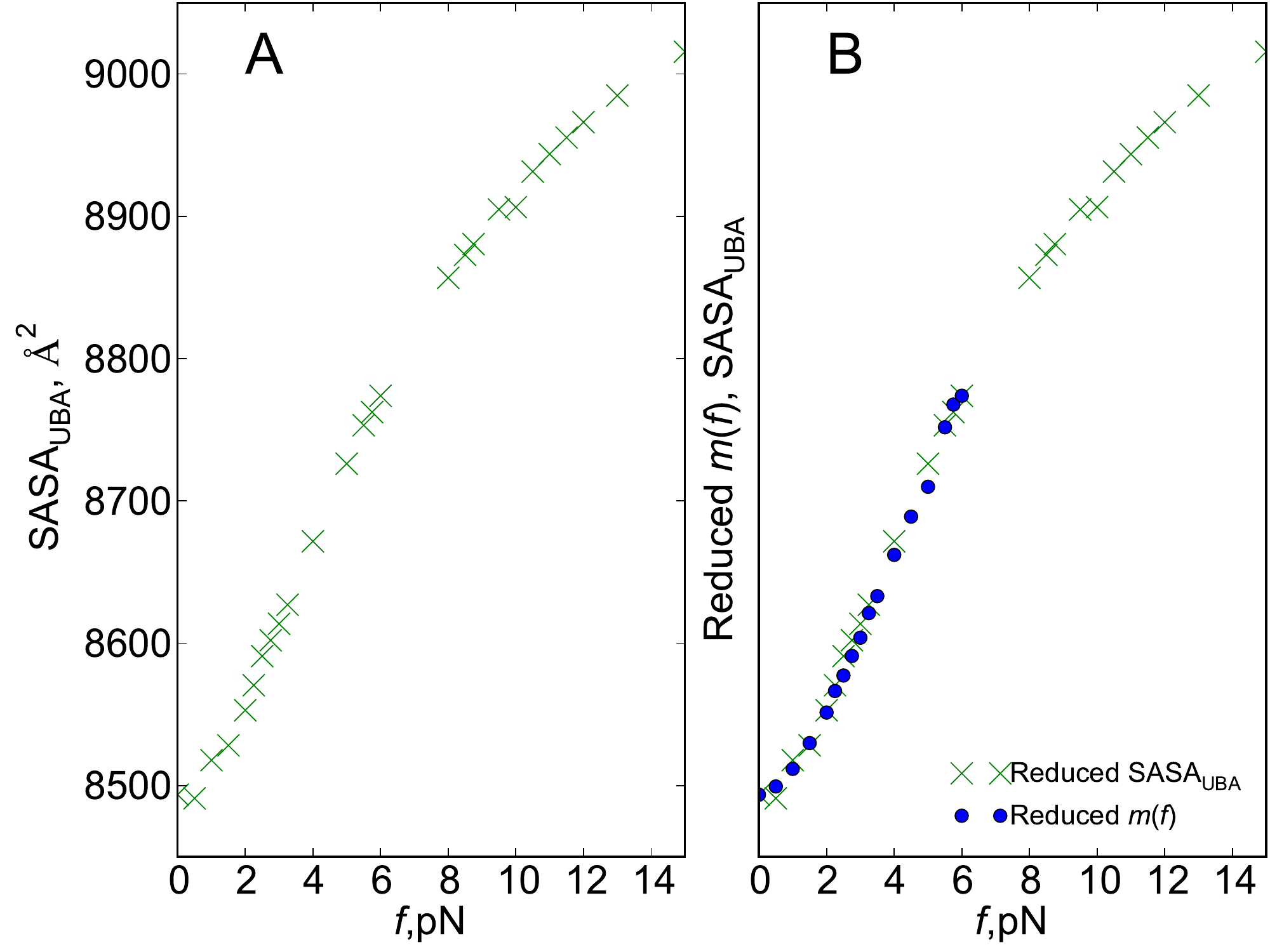}
       \caption{(A) The dependence of the solvent accessible surface area (SASA) in the unfolded ensemble on $f$ at $T=340$ K and $[C]=0$. (B) SASA in the unfolded ensemble dependence on $f$ overlayed with the $m(f)$, where $m(f)$  in $\Delta G_\mathrm{NBA}([C],T,f)=\Delta G_\mathrm{NBA}([0],T,f)+m(f)[C]$, is obtained from linear regression of $\Delta G_\mathrm{NBA}([C])$ shown in Fig.1C of the main text. The precise coincidence shows that all of the dependence of $m(f)$ on $f$ comes from changes in the SASA in the unfolded ensemble as $[C]$ increases.
\label{fig:mvssasa}}
\end{figure}


\begin{figure}
	\centering
		\includegraphics[width=8.7cm]{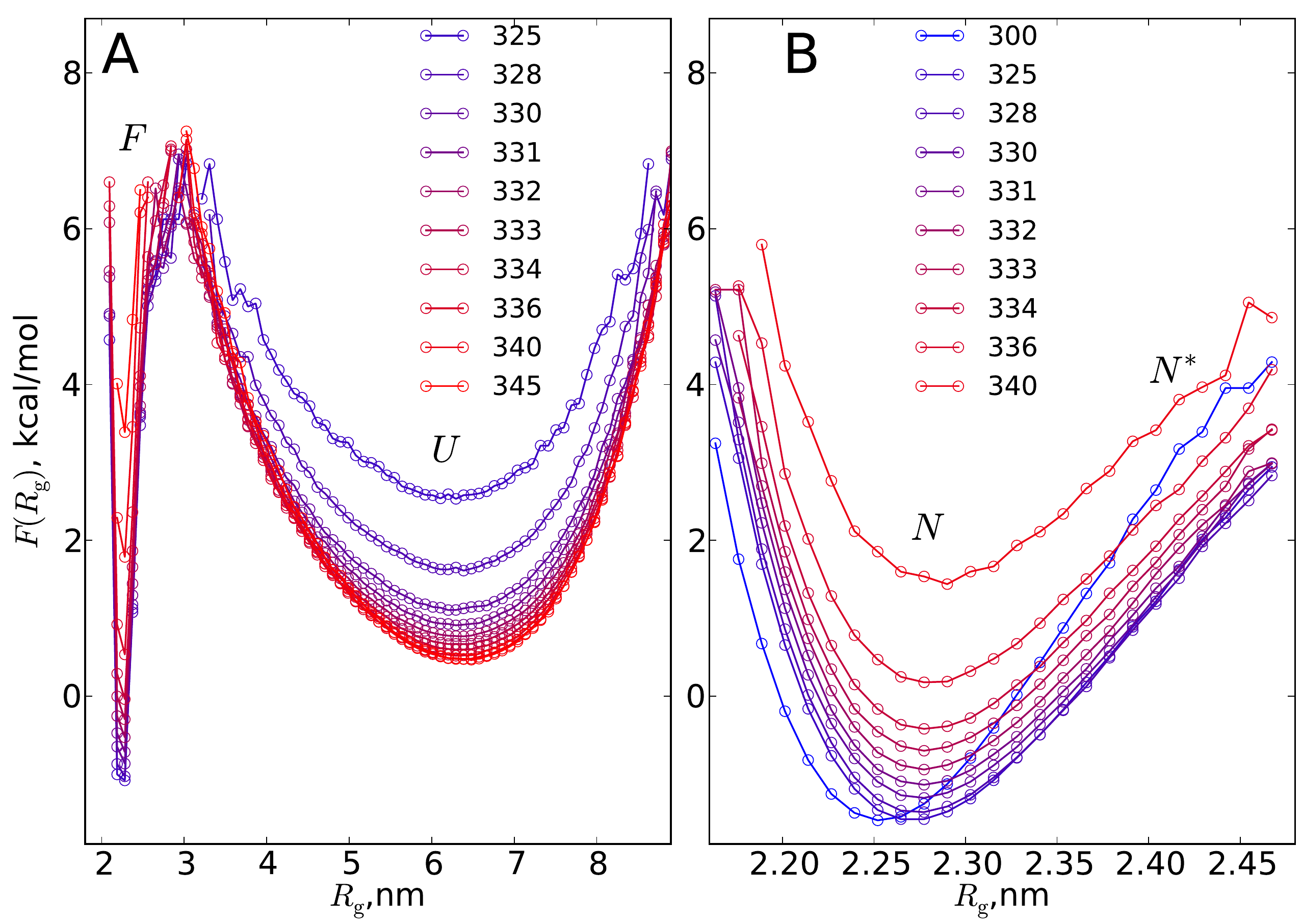}
       \caption{Free energy profiles at $f=7.5$ pN. (A) $F(R_g)$ at various temperatures. (B) Even when zooming into the $R_g$ range $(2.2 < R_g < 2.45)$ nm, the $N^*$ state in the fine structure in $F(R_\mathrm{g})$ is not visible unlike in the free energy profiles as a function of the end-to-end distance (see Fig. 3 in the main text).
\label{fig:reehistrg}}
\end{figure}

\begin{figure}
	\centering
		\includegraphics[width=8.7cm]{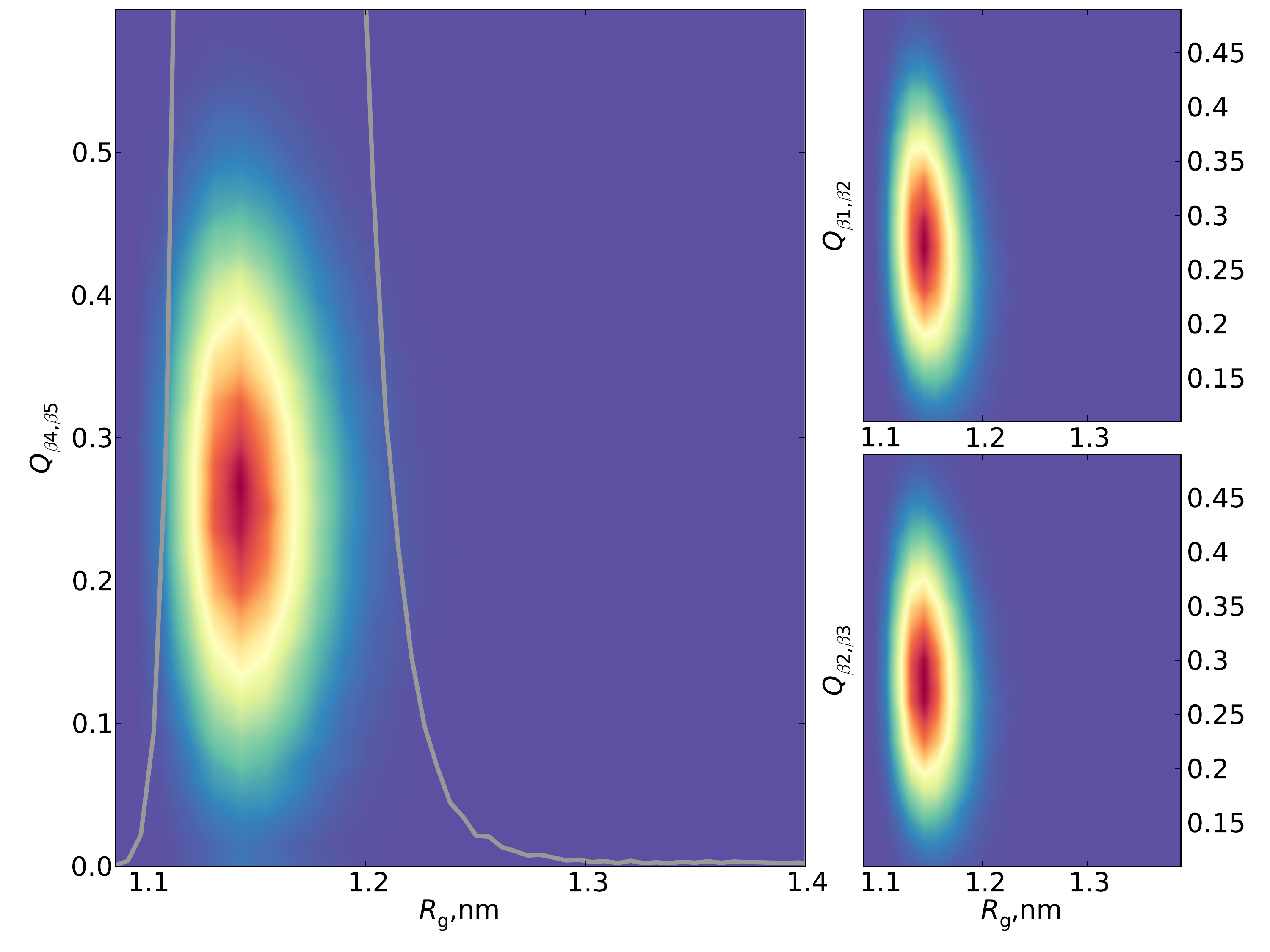}
       \caption{$N^*$ state is hidden in the ($R_\mathrm{g},Q_{x,y}$) histograms expressed in terms of $R_g$ and contacts between various secondary structural elements.
\label{fig:rghistqb4b5}}
\end{figure}


\begin{figure}
	\centering
		\includegraphics[width=8.7cm]{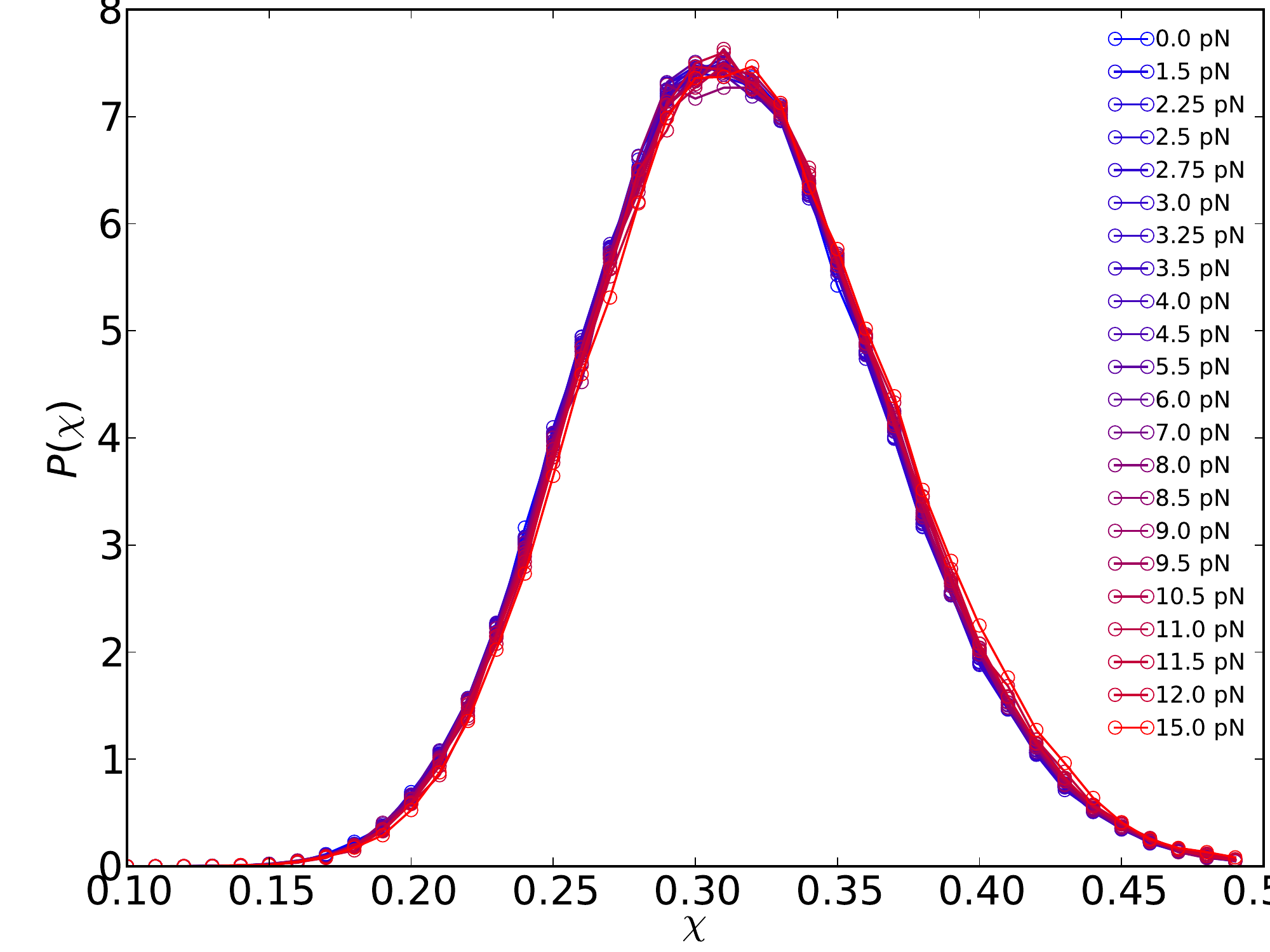}
       \caption{Probability density of the structural overlap parameter $\chi$ in the NBA. The $N^*$ state, which is part of the NBA,  is not discernible in $\chi$ at any force value. 
\label{fig:fshistchi}}
\end{figure}

\begin{figure}
	\centering
		\includegraphics[width=8.7cm]{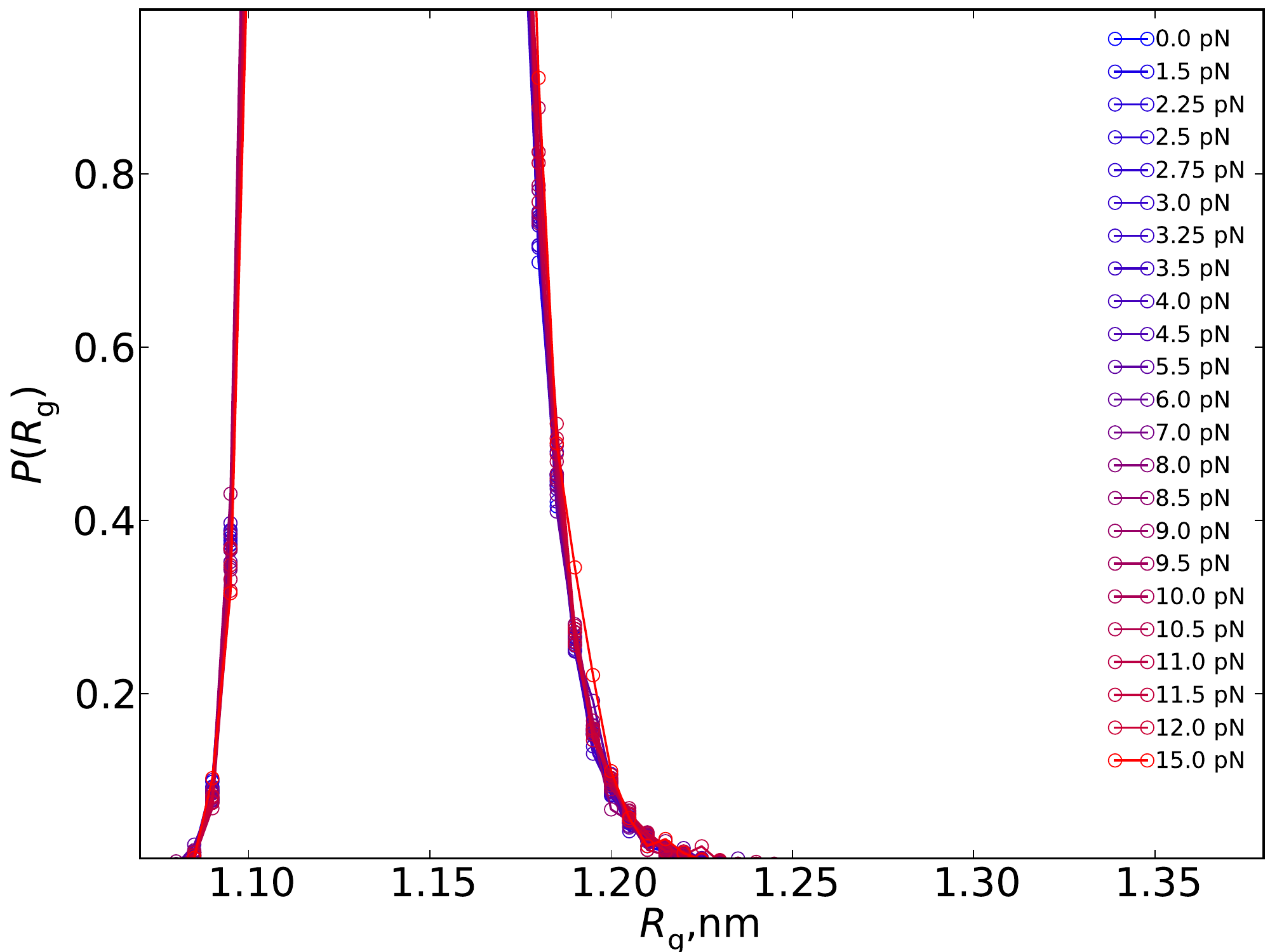}
       \caption{Zoom into the NBA as a probability density of $R_\mathrm{g}$. The $N^*$ state is not discernible. Figures S3-S6 show that a suitable order parameter (end-to-end distance) is needed to characterize the $N^*$ state.
\label{fig:fstatehistrg}}
\end{figure}

\begin{figure}
	\centering
		\includegraphics[width=8.7cm]{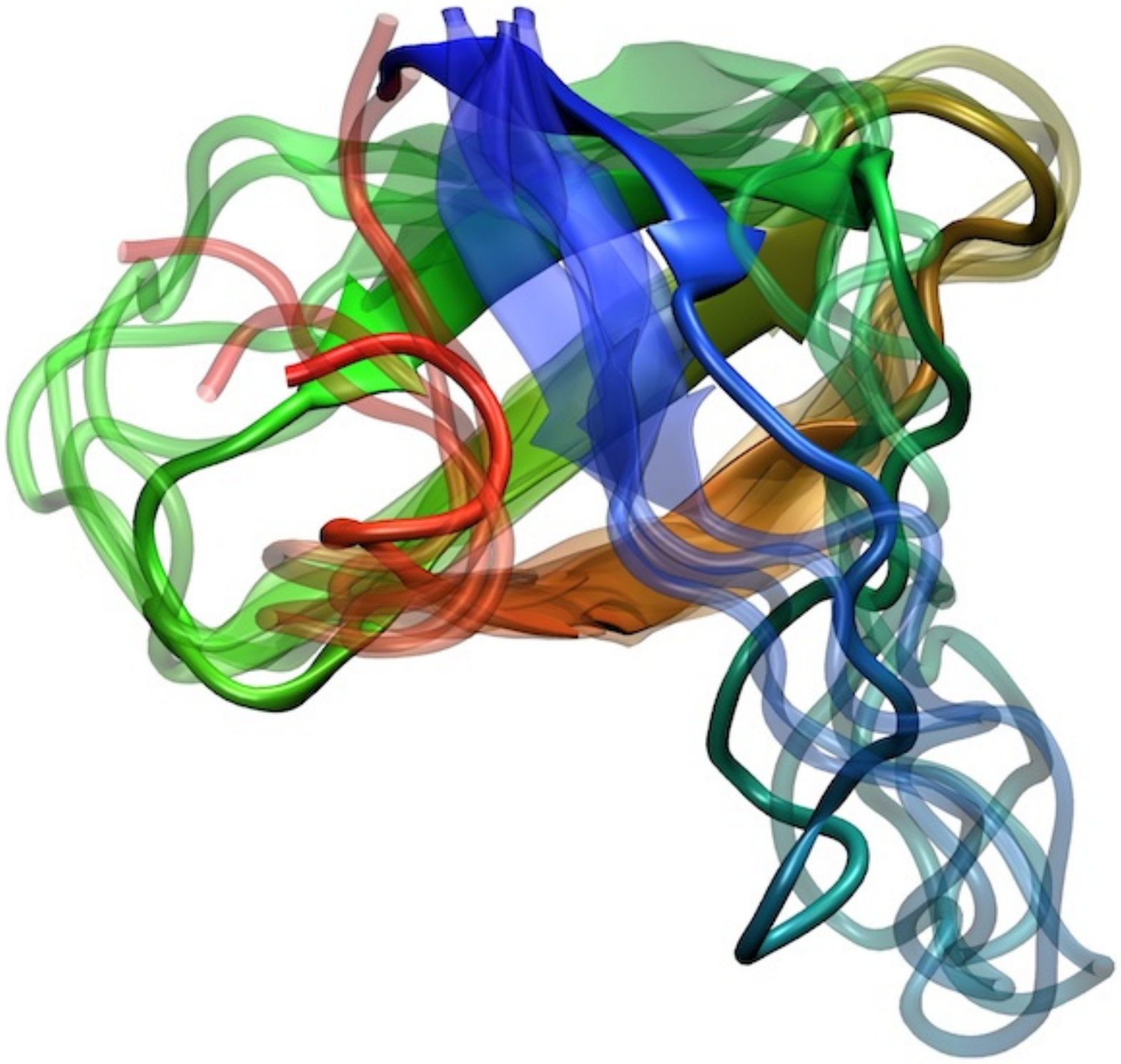}
       \caption{The $N^*$ conformation of the Fyn SH3 domain from NMR study by Kay and coworkers\cite{Neudecker:2012kn} (solid) overlayed with several different $N^*$ snapshots from our simulations. The structures only differ from the native state in the $\beta 5$ (C-terminal) strand which is unstructured in $N^*$ and not resolved in the NMR. The average RMSD between the Fyn SH3 $N^*$ and the src SH3 $N^*$ is 3.3 $\AA$.
\label{fig:rmsdtofyn}}
\end{figure}


\begin{figure}
	\centering
		\includegraphics[width=8.7cm]{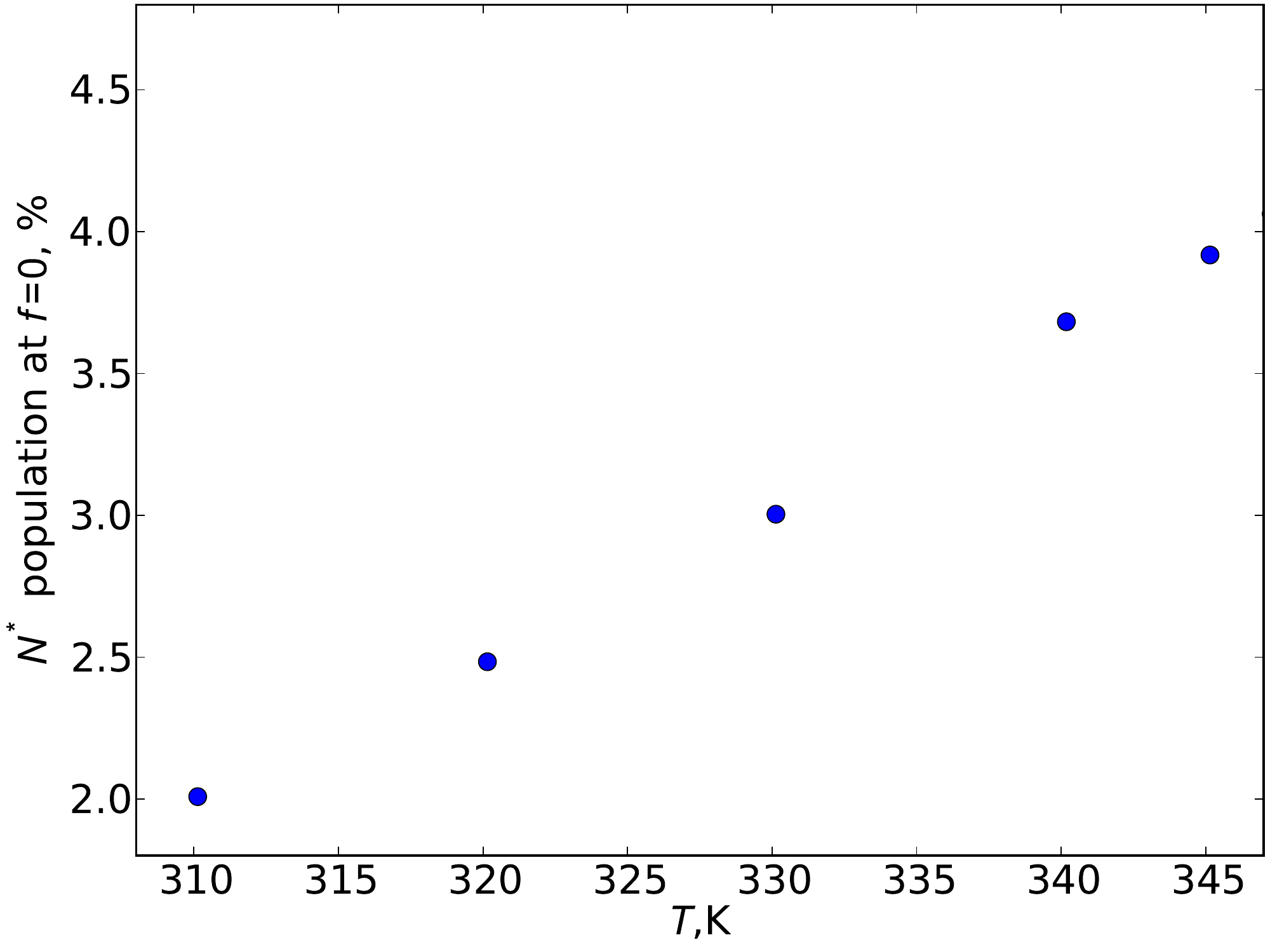}
       \caption{Population of the $N^*$ state at $f=0$ at different temperatures obtained from free energy profiles from simulations at $f=0$. 
\label{fig:nstarpopf0}}
\end{figure}


\begin{figure}
	\centering
		\includegraphics[width=8.7cm]{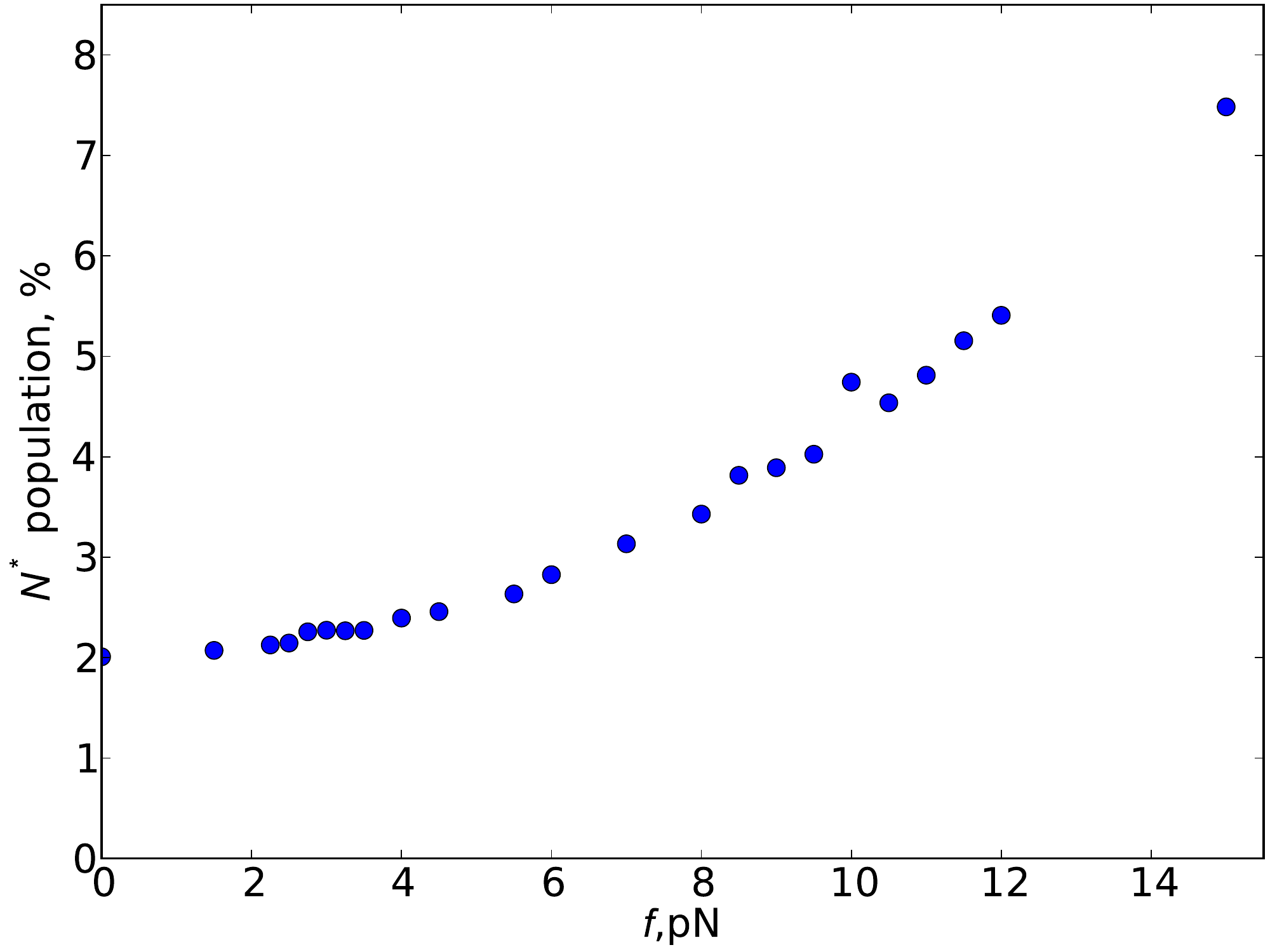}
       \caption{Population of the $N^*$ state at different forces calculated from the histograms given in Fig.3 of the main text. 
\label{fig:nstarpop}}
\end{figure}

\end{document}